\documentclass[iop,numberedappendix]{emulateapj}

\pdfoutput=1

\usepackage{amsmath,natbib,graphicx}
\usepackage{epsf}
\usepackage{epstopdf}
\input{epsf}
\usepackage{epsfig}
\usepackage{color}
\DeclareGraphicsExtensions{.jpg,.pdf,.png,.eps,.ps}

\newcommand{\ltsima}{$\; \buildrel < \over \sim \;$}
\newcommand{\ltsim}{\lower.5ex\hbox{\ltsima}}
\newcommand{\um}{$\mu \mathrm{m}$~}
\newcommand{\beq}{\begin{equation}}
\newcommand{\eeq}{\end{equation}}

\newcommand{\srcfull}{SPT-S\,J053816$-$5030.8}
\newcommand{\smg}{SPT\,0538$-$50}

\def\msol   {\ifmmode{{\rm M}_{\odot} }\else{M$_{\odot}$}\fi}
\def\lsol   {\ifmmode{{\rm L}_{\odot}}\else{${\rm L}_{\odot}$}\fi}
\def\ts     {\thinspace}
\def\kms    {\ifmmode{{\rm \ts km\ts s}^{-1}}\else{\ts km\ts s$^{-1}$}\fi}

\def\ctwo {\ifmmode{{\rm C}{\rm \small I}(^3\!P_2\!\to^3\!P_1)}
     \else{C\ts {\scriptsize I}{\small$(^3\!P_2\!\to^3\!\!\!P_1)$}}\fi}
     
\def\coone{{$^{12}$CO(1$\to$0)}}

\def\coseven{{$^{12}$CO(7$\to$6)}}
\def\coeight{{$^{12}$CO(8$\to$7)}}

\def\Cambridge{1}
\def\Arizona{2}
\def\UPenn{3}
\def\Dal{4}
\def\Caltech{5}
\def\CfA{6}
\def\ESO{7}
\def\Diego{8}
\def\KICPChicago{9}
\def\EFIChicago{10}
\def\JPL{11}
\def\Miss{12}
\def\PhysicsUChicago{13}
\def\AAUChicago{14}
\def\ANL{15}
\def\Davis{16}
\def\UFlorida{17}
\def\UCL{18}
\def\McGill{19}
\def\Berkeley{20}

\def\Catolica{22}
\def\IfA{23}
\def\Colorado{24}
\def\UCLA{25}
\def\ATNF{26}
\def\Carnegie{27}
\def\MPIfR{28}
\def\APC{29}

\begin{document}

\title{\smg: Physical Conditions in the ISM of a strongly lensed dusty star-forming galaxy at z=2.8}

\shorttitle{}



\shorttitle{Physical conditions within a lensed SMG at $z=2.8$}
\shortauthors{SPT collaboration}

\author{
M.~S.~Bothwell$^{\Cambridge,\Arizona}$,
J.~E.~Aguirre$^{\UPenn}$,
S.~C.~Chapman$^{\Dal}$,
D.~P.~Marrone$^{\Arizona}$,   
J.~D.~Vieira$^{\Caltech}$,
M.~L.~N.~Ashby$^{\CfA}$,
M.~Aravena$^{\ESO,\Diego}$,
B.~A.~Benson$^{\KICPChicago,\EFIChicago}$, 
J.~J.~Bock$^{\Caltech,\JPL}$,
C.~M.~Bradford$^{\JPL}$,
M.~Brodwin$^{\Miss}$,
J.~E.~Carlstrom$^{\KICPChicago,\PhysicsUChicago,\EFIChicago,\AAUChicago,\ANL}$, 
T.~M.~Crawford$^{\KICPChicago,\AAUChicago}$, 
C.~de~Breuck$^{\ESO}$,
T.~P.~Downes$^{\Caltech}$,
C.~D.~Fassnacht$^{\Davis}$,
A.~H.~Gonzalez$^{\UFlorida}$, 
T.~R.~Greve$^{\UCL}$,	
B.~Gullberg$^{\ESO}$, 
Y.~Hezaveh$^{\McGill}$,
G.~P.~Holder$^{\McGill}$, 
W.~L.~Holzapfel$^{\Berkeley}$, 
E.~Ibar$^{\Catolica}$, 
R.~Ivison$^{\IfA}$, 
J.~Kamenetzky$^{\Colorado}$, 
R.~Keisler$^{\KICPChicago,\PhysicsUChicago}$, 
R.~E.~Lupu$^{\UPenn}$,
J.~Ma$^{\UFlorida}$, 
M.~Malkan$^{\UCLA}$,
V.~McIntyre$^{\ATNF}$,
E.~J.~Murphy$^{\Carnegie}$,
H.~T.~Nguyen$^{\JPL}$,
C.~L.~Reichardt$^{\Berkeley}$, 
 M.~Rosenman$^{\UPenn}$,
J.~S.~Spilker$^{\Arizona}$,
B.~Stalder$^{\CfA}$, 
A.~A.~Stark$^{\CfA}$, 
M.~Strandet$^{\MPIfR}$, 
J.~Vernet$^{\ESO}$, 
A.~Wei\ss$^{\MPIfR}$,
N.~Welikala$^{\APC}$
}
\altaffiltext{\Cambridge}{Cavendish Laboratory, University of Cambridge, JJ Thompson Ave, Cambridge CB3 0HA, UK}
\altaffiltext{\Arizona}{Steward Observatory, University of Arizona, 933 North Cherry Avenue, Tucson, AZ 85721, USA}
\altaffiltext{\UPenn}{University of Pennsylvania, 209 South 33rd Street, Philadelphia, PA 19104, USA}
\altaffiltext{\Dal}{Dalhousie University, Halifax, Nova Scotia, Canada}
\altaffiltext{\Caltech}{California Institute of Technology, 1200 E. California Blvd., Pasadena, CA 91125, USA}
\altaffiltext{\CfA}{Harvard-Smithsonian Center for Astrophysics, 60 Garden Street, Cambridge, MA 02138, USA}
\altaffiltext{\ESO}{European Southern Observatory, , Alonso de Cordova 3107, Casilla 19001 Vitacura Santiago, Chile.}
\altaffiltext{\Diego}{Universidad Diego Portales, Facultad de Ingenier\'{\i}a, Av. Ej\'{e}rcito 441, Santiago, Chile}
\altaffiltext{\KICPChicago}{Kavli Institute for Cosmological Physics, University of Chicago, 5640 South Ellis Avenue, Chicago, IL 60637, USA}
\altaffiltext{\EFIChicago}{Enrico Fermi Institute, University of Chicago, 5640 South Ellis Avenue, Chicago, IL 60637, USA}
\altaffiltext{\JPL}{Jet Propulsion Laboratory, 4800 Oak Grove Drive, Pasadena, CA 91109, USA}
\altaffiltext{\Miss}{Department of Physics and Astronomy, University of Missouri, 5110 Rockhill Road, Kansas City, MO 64110, USA}
\altaffiltext{\PhysicsUChicago}{Department of Physics, University of Chicago, 5640 South Ellis Avenue, Chicago, IL 60637, USA}
\altaffiltext{\AAUChicago}{Department of Astronomy and Astrophysics, University of Chicago, 5640 South Ellis Avenue, Chicago, IL 60637, USA}
\altaffiltext{\ANL}{Argonne National Laboratory, 9700 S. Cass Avenue, Argonne, IL, USA 60439, USA}
\altaffiltext{\Davis}{Department of Physics,  University of California, One Shields Avenue, Davis, CA 95616, USA}
\altaffiltext{\UFlorida}{Department of Astronomy, University of Florida, Gainesville, FL 32611, USA}
\altaffiltext{\UCL}{Department of Physics and Astronomy, University College London, Gower Street, London WC1E 6BT, UK}
\altaffiltext{\McGill}{Department of Physics, McGill University, 3600 Rue University, Montreal, Quebec H3A 2T8, Canada}
\altaffiltext{\Berkeley}{Department of Physics, University of California, Berkeley, CA 94720, USA}
\altaffiltext{\Catolica}{Instituto de Astrof\'isica, Facultad de F\'isica, Pontificia Universidad Cat\'olica de Chile, Casilla 306, Santiago 22, Chile}
\altaffiltext{\IfA}{Institute for Astronomy, Blackford Hill, Edinburgh EH9 3HJ, UK}
\altaffiltext{\Colorado}{Center for Astrophysics and Space Astronomy, University of Colorado, 389 UCB, Boulder, Colorado, 80309, USA}
\altaffiltext{\UCLA}{Department of Physics and Astronomy, University of California, Los Angeles, CA 90095-1547, USA}
\altaffiltext{\ATNF}{Australia Telescope National Facility, CSIRO, Epping, NSW 1710, Australia}
\altaffiltext{\Carnegie}{Observatories of the Carnegie Institution for Science, 813 Santa Barbara Street, Pasadena, CA 91101, USA}
\altaffiltext{\MPIfR}{Max-Planck-Institut f\"{u}r Radioastronomie, Auf dem H\"{u}gel 69 D-53121 Bonn, Germany}
\altaffiltext{\APC}{AstroParticule et Cosmologie, Universit\'{e} Paris Diderot, CNRS/IN2P3, CEA/lrfu, Observatoire de Paris, Sorbonne Paris Cit\'{e}, 10, rue Alice Domon et L\'{e}onie Duquet, 75205 Paris Cedex 13, France}





\begin{abstract}
We present observations of \srcfull, a gravitationally-lensed dusty star forming galaxy (DSFG) at $z=2.7817$, first discovered at millimeter wavelengths by the South Pole Telescope. \smg\ is typical of the brightest sources found by wide-field millimeter-wavelength surveys, being lensed by an intervening galaxy at moderate redshift (in this instance, at $z=0.441$). We present a wide array of multi-wavelength spectroscopic and photometric data on \smg, including 
data from ALMA, \textit{Herschel} PACS and SPIRE, \textit{Hubble}, \textit{Spitzer}, VLT, ATCA, APEX, and the SMA. 
We use high resolution imaging from HST to de-blend \smg, separating DSFG emission from that of the foreground lens. Combined with a source model derived from ALMA imaging (which suggests a magnification factor of $21 \pm 4$), we derive the intrinsic properties of \smg, including the stellar mass, far-IR luminosity, star formation rate, molecular gas mass, and -- using molecular line fluxes -- the excitation conditions within the ISM.  
The derived physical properties argue that we are witnessing compact, merger-driven star formation in \smg, similar to local starburst galaxies, and unlike that seen in some other DSFGs at this epoch.  


\end{abstract}

\section{Introduction}\label{sec:intro}

The population of high-redshift, luminous dusty starbursts known as `submillimeter galaxies' was first discovered in 1998, with the SCUBA instrument on the JCMT \citep{holland99}. Since then, a picture has emerged of a population of massive (M$_* \sim 10^{11}$ M$_{\sun}$; \citealt{hainline11, michalowski12}) and gas-rich (M$_{\rm gas}$ $>10^{10}$ M$_{\sun}$; \citealt{frayer98,neri03,greve05,bothwell13}) galaxies, with IR luminosities that imply prodigious star formation rates (100s to $>$1000 ${\rm M}_{\sun} {\rm yr}^{-1}$; \citealt{kovacs06}). It is clear that they represent a population of significant astrophysical importance, with properties that remain a considerable challenge for conventional galaxy evolution models \citep{baugh05, lacey10, benson12b, hayward12}.


The advent of millimeter/submillimeter (submm) instruments capable of surveying large areas of sky, including \textit{Herschel}-SPIRE \citep{pilbratt10, griffin10} and the South Pole Telescope (SPT; \citealt{carlstrom11}), has opened new vistas for the study of these submm bright systems (\citealt{vieira10}; \citealt{negrello10}). The brightest of these submm sources, with $S_{\rm 850 \mu m}>100$mJy, have such low sky density ($< 1 \; {\rm deg}^{-2}$) that large area surveys are the only way to detect them in statistically significant numbers. Follow-up imaging in the submm at high spatial resolution ($<1\arcsec$) has confirmed the initial predictions that these objects represent a population of lensed, dusty star-forming galaxies (DSFGs) magnified by gravitational lensing \citep{negrello10,vieira13,hezaveh13}. %

In recent years, the study of lensed starburst systems at high redshift has offered invaluable insight into the behavior and physics of these dust-obscured systems. Strong gravitational lensing provides an effective boost in both sensitivity and angular resolution, allowing highly detailed analyses of lensed systems that would not be possible in their unlensed counterparts. 

The $z=2.3$ galaxy SMM J2135$-$0102 -- lensed by a factor of $\sim 30$ -- has been extensively studied, with the lensing magnification allowing detailed observations of a large rotating molecular disc \citep{swinbank11}, a dusty IR spectral energy distribution (SED) \citep{ivison10}, and a well-constrained CO spectral line energy distribution (SLED) implying ISM excitation conditions similar to local starbursts \citep{danielson11}. \cite{fu12} present a comprehensive analysis of a H-ATLAS-selected lensed DSFG at $z=3.2$, a system for which the lensing magnification has revealed a highly complex, clumpy structure, and an obscured UV-to-far IR SED. In addition, \cite{wardlow12} present 13 candidate gravitationally-lensed DSFGs from the Herschel Multi-tiered Extragalactic Survey (HERMES), finding that their intrinsic sub-mm fluxes typically lie below the Herschel selection limit, and as such represent systems which can only be studied via gravitational lensing. 


\srcfull\ (hereafter \smg) is the first lensed DSFG selected by the SPT survey to be characterized in detail. \smg\ is taken from the catalog of sources from the first 87 deg$^2$ in the SPT survey \citep{vieira10}, with single-dish submm followup presented in \citep{greve12}. It was targeted for imaging with the Atacama Large Millimeter/submillimeter Array (ALMA) -- these results, which were used to construct a physically motivated lens model of the system, were presented by \cite{hezaveh13}. With this lens model in hand we present a range of multi-wavelength observations of the source, and derive the intrinsic luminosity, star formation rate, stellar mass, molecular gas mass, and physical conditions of this interesting galaxy. 

The paper is organized as follows: in \S\ref{sec:obs} we present our multiwavelength observations of \smg, which we divide into imaging (\S2.1) and spectroscopy (\S2.2). In \S3 we present our analysis, and in \S4 we discuss the implications for the physical properties of \smg\ . We conclude in \S5. All magnitudes in this work are given in the Vega system. Throughout this work we adopt a cosmological model with ($\Omega_m, \; \Omega_{\Lambda},\; \mathrm{H}_0) = (0.27,\; 0.73,\; 71$ km\,s$^{-1}$\,Mpc$^{-1}$), and a \cite{kroupa01} initial mass function  (which is very similar to the \citealt{chabrier03} IMF; masses and luminosities derived using a Kroupa IMF can be converted to Chabrier by multiplying by a factor 1.03).

\section{Observations and results}\label{sec:obs}

Since its discovery, \smg\ (ALMA co-ordinates RA: 05:38:16.83, DEC: 50:30:52.0) has been observed in a multitude of wavebands, from visible to radio wavelengths. The resulting datasets enable us to investigate the intrinsic physical properties of the interstellar medium (ISM), while also constraining the stellar mass. Here we present the observational data collected for \smg\, starting with submillimeter and millimeter imaging data, moving on to NIR and optical imaging, and finally millimeter-wavelength spectroscopy.

\subsection{Imaging}

\begin{figure}
\centering
{\includegraphics[width=8cm]{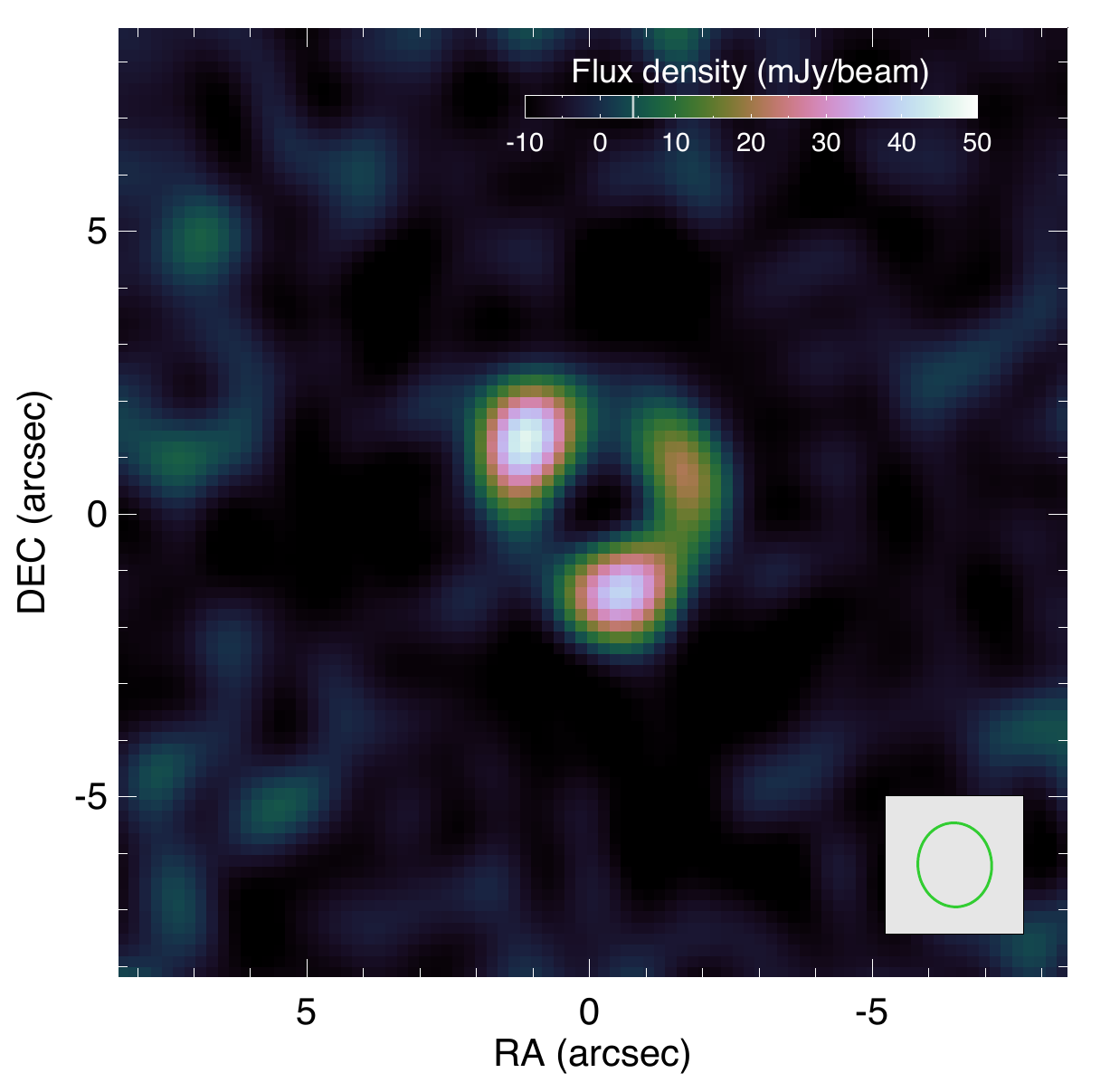}}
\caption{ALMA Band 7 (= 349.8 GHz) continuum image of \smg\, taken with an integration time of 60 seconds. The synthesized beam FWHM ($1.3'' \times 1.5''$, at PA = 5$^{\circ}$) is shown inset in the lower right. The observations resolve the submillimeter emission into a clear Einstein ring with a radius of $1.8''$, conclusively demonstrating the lensed origin of \smg.}
\label{fig:ALMA}
\end{figure}

Here we present observational details for the imaging data used in this work. Data from the SPT (\S2.1.1), ALMA (\S2.1.2), ATCA (\S2.1.3) and APEX (\S2.1.4) have been previously published by our group (see the relevant sections for references). We also present unpublished data from Herschel (\S2.1.5), Spitzer/IRAC (\S2.1.6), VLT/ISAAC (\S2.1.7) and HST/WFC3 (\S2.1.8).

\subsubsection{SPT}
\label{sec:alma}

\smg\ was initially detected at 1.4mm and 2.0mm wavelengths by the South Pole Telescope. Details of these observations are given by \cite{vieira10}. An updated description of the SPT survey and point source catalog can be found in \citet{mocanu13}.

\subsubsection{ALMA}
\label{sec:alma}

The ALMA observations are described in \citet{hezaveh13} -- here we provide a summary.
\smg\ was imaged by ALMA in Cycle 0 (2011.0.00958.S; PI: D. Marrone) on 2011 November 16 for a total on-source integration time of 60 seconds.  Sixteen antennas were incorporated into the array at the time of observation.  
Baselines extended from 20--150k$\lambda$, creating a beam with a full width at half maximum (FWHM) of $\sim1.4''$. 
The Band 7 local oscillator was tuned to 343.8~GHz, with four correlator spectral windows configured to receive 1.875~GHz wide bands 
in two polarizations each at central frequencies of 336.8, 338.675, 348.925, and 350.8~GHz, for a total effective bandwidth of 7.5GHz.  
Bandpass calibration was performed using observations of quasar J0538-440, while complex gain solutions were calculated using quasar J0519-4546.  Absolute flux calibration was determined using observations of Callisto. 

To register the ALMA data with images at other wavelengths, the primary phase calibrator was chosen to be registered to the 
International Celestial Reference Frame \citep[ICRF;][]{Ma98}. The calibrator and an additional test source, J0540-5418, are both 
registered to the ICRF with precision of a few milliarcseconds \citep{petrov11}. This test source was observed less than three minutes 
after \smg\ to verify the astrometric calibration. Transferring phase and amplitude calibration from the phase calibrator J0519-4546, 9.1$^\circ$ away, 
recovers the position of the test source to within $0.09''$, suggesting a similar precision for \smg, which is 5.6$^\circ$ from the calibrator and 3.8$^\circ$ from the test source. 

Our ALMA Band 7 compact-array observations (see Fig. \ref{fig:ALMA}) have an rms noise of 1.0 mJy/beam. \smg\ is detected with a S/N ratio of $\sim 50$,  and is resolved into a clear Einstein ring and conclusively indicating a lensed origin for the source.

\subsubsection{ATCA}
\label{sec:atca}

\smg\ was observed with the Australia Telescope Compact Array (ATCA) on 2012 July 31, and was well detected in both \coone and the 35 GHz continuum. The final data resulted in a continuum level of $130 \pm 20 \; \mu$Jy, a continuum RMS of 20 $\mu$Jy at 35 GHz, and an RMS of 0.60 mJy beam$^{-1}$ per 2 MHz channel at 30.6 GHz (near the location of the CO emission line). Further details are reported in \cite{aravena13}.

\subsubsection{APEX LABOCA \& SABOCA}
 \smg\ was imaged with the Atacama Pathfinder Experiment (APEX) Submillimetre APEX Bolometer Camera (SABOCA) and the Large APEX Bolometer Camera (LABOCA) instruments in May of 2010, in excellent weather conditions (see \citealt{siringo09} for instrument details). On-source integration times for SABOCA 350$\mu$m and LABOCA 870$\mu$m observations were approximately 3 hours and 1 hour respectively, reaching respective noise levels of 30 mJy/beam, and 6 mJy/beam. \smg\ was detected with high significance in both bands, with resulting fluxes S$_{350\mu m} = 336 \pm 88$ mJy, and S$_{870\mu m} = 125 \pm 7$ mJy (see \cite{greve12} for further details on these observations). \smg\ is unresolved in both bands, which have beam FWHMs of $\sim 20'' \; ({\rm at} \; {870\mu}$m) and $\sim 8'' \; ({\rm at} \; {350\mu }$m).

\subsubsection{Herschel PACS \& SPIRE}
The acquisition and reduction of data from both PACS
and SPIRE \citep{griffin10}
are described extensively in 
\citet{clements10}, \citet{ibar10}, and \cite{rigby11}.

The PACS data were acquired in programs OT1\_jvieira\_4 and OT1\_dmarrone\_1 using approximately orthogonal scans centered on the target at medium speed (i.e., with the telescope tracking at $20''$ s$^{-1}$) using `scan map' mode, spending a total of 180\,s on source per program. Data were recorded simultaneously at 100 and 160\,$\mu$m. Each scan comprises ten separate $3'$ strips, each offset orthogonally by $4''$. The data were then handled by a variant of the reduction pipeline presented in \cite{ibar10}. The resulting noise levels were $\sigma\approx 4$ and 7\,mJy at 100 and
160\,$\mu$m, respectively (calculated using random aperture photometry).

The SPIRE data from program OT1\_jvieira\_4 consist of a triple
repetition map, with coverage complete to a radius of 5 arcmin from
the nominal SPT position. The maps were produced via the standard
reduction pipeline HIPE v9.0, the SPIRE Photometer Interactive
Analysis (SPIA) package v1.7, and the calibration product v8.1.
Photometry was extracted by fitting a Gaussian profile to the position
of the SPT detection and the noise was estimated by taking the RMS in
the central 5 arcmin of the map which is then added in quadrature to
the absolute calibration uncertainty.


\subsubsection{Spitzer IRAC}

\smg\ was also observed at 3.6$\mu$m and $4.5\mu$m by the Infrared Array Camera (IRAC; \citealt{fazio04}) on board the {\sl Spitzer Space Telescope} \citep{werner04} on 2009 August 2, as part of a large program to obtain follow-up imaging of the SPT sources (PID 60194; PI Vieira). The basic calibrated data, pre-processed by the Spitzer Science Center (SSC)'s standard pipeline, were combined into a re-sampled mosaic image using IRACproc \citep{schuster06} as a wrapper to optimize the performance of the MOPEX software package (Makovoz \& Marleau 2005). The resultant IRAC mosaic images
 have an angular resolution of $\sim 2''$, and a re-sampled pixel scale of $0.6''$/pixel. \

The foreground lens galaxy is well detected in both IRAC bands, with flux densities at 3.6$\mu$m and $4.5\mu$m of $S_{3.6} = 0.38 \pm 0.01$ mJy and $S_{4.5} =0.40 \pm 0.01$ mJy.
Due to the small angular separation between the lensed image of \smg\ and the foreground lens, the (relatively) large IRAC beam causes emission from the two sources to be blended. In \S\ref{sec:results}, we describe our de-blending method which allows us to separate emission from \smg\ from that of the foreground lens. 

\subsubsection{VLT ISAAC}

A deep $Ks$-band image of  \smg\  was obtained using
ISAAC (Moorwood et al. 1998), mounted on UT1 (Antu) of VLT at the
European Southern Observatory (ESO) in Paranal, Chile. ISAAC is equipped
with a $1024 \times 1024$ pixel Hawaii Rockwell array, with a pixel scale of
0.148$''$ px$^{-1}$, giving a field of view of $\sim150\times150$
arcsec$^2$. The observations were performed in service mode under
photometric conditions on 2010 March 6. The seeing, as derived from the
FWHM size of stars in each frame, was mostly excellent during the
observations, ranging from $0.4''$ to $0.6''$. Note that at the redshift
of \smg, observations in the Ks-band probe
rest-frame $\sim$5800\AA, or roughly the R-band.

The images were secured using individual exposures of 2 minutes per
frame, and a jitter procedure which produces a set of frames at randomly
offset telescope positions within a box of $10\times10 \arcsec$ . The
total integration time was 30 minutes per target. 
Data reduction was performed using the ESO pipeline for jitter imaging data
(Devillard 2001). Each frame was dark-subtracted and flat-fielded by
normalized flat field obtained from twilight sky images. Sky subtraction
was performed using median averaged and scaled sky frames obtained
combining jittered exposures of the same field. Sky-subtracted images
were aligned to sub-pixel accuracy, and co-added. Photometric
calibration was performed from comparison with 2MASS magnitudes of
bright stars available in the field. The estimated internal photometric
accuracy is $\sim$0.1 mag.

\subsubsection{HST WFC3-IR}

 \smg\ was observed with the Hubble Space Telescope Wide Field Camera 3 on  2011 October 10, as part of program 12659 (PI: Vieira). We use the F110W and F160W filters, the reddest high-throughput filters available. One complete orbit was dedicated to this source, split evenly between the two filters. 
The data presented here are based on data  products delivered by the HST pipeline, and photometry is defined following the WFC3 handbook.

\subsection{Spectroscopy}

Here we present observational details for the spectroscopic data used in this work. The VLT spectrum (\S2.2.4) appeared previously in \cite{hezaveh13} -- all other data are reported here for the first time.

\subsubsection{Z-Spec Millimeter Spectroscopy}
\label{sec:zspec}

Z-Spec is a broadband, millimeter-wave direct-detection grating spectrometer operating from 185 to 305 GHz.  The channel resolution, which is fixed by the spectrometer geometry, ranges from $R=250$ to $R=400$ (700 km s$^{-1}$ to 1100 km s$^{-1}$).  Further details about the Z-Spec instrument may be found in \citet{naylor03}.

\begin{figure}
\centering
{\includegraphics[width=8.7cm]{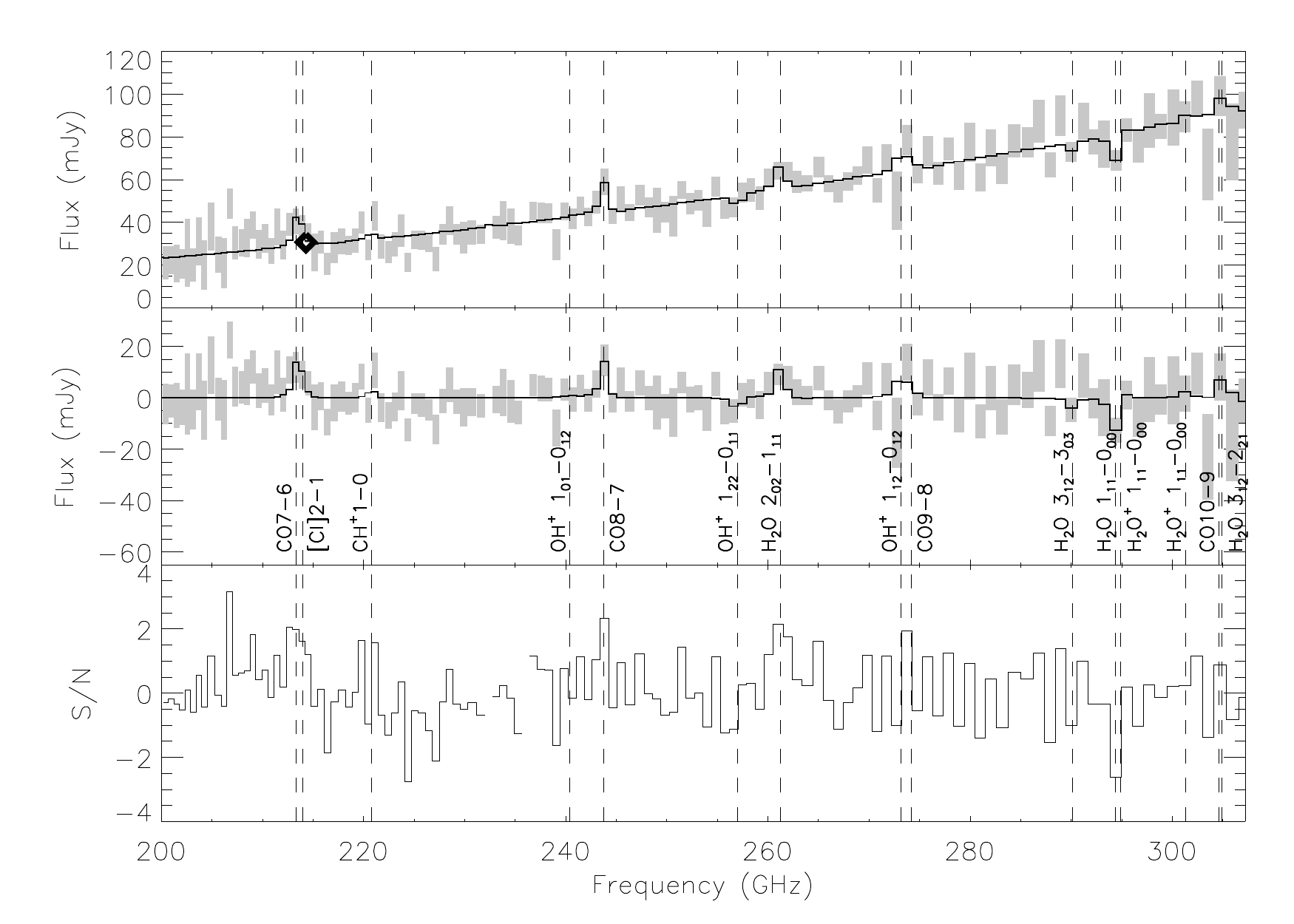}}
\caption{APEX/Z-Spec spectrum of \smg. The top panel shows the spectrum with continuum, while the central panel shows the continuum-subtracted spectrum. The lower panel shows the S/N. The model spectrum (including all emission/absorption lines) is shown as a black overlay, and all lines are identified and highlighted. See description in \S\ref{sec:zspec}. The expected frequencies of other molecular lines lying within the band are also marked for reference. The redshift of \smg\ was originally obtained with this Z-Spec spectrum.}
\label{fig:zspec-spectrum}
\end{figure}

Observations of SPT 0538-50 were obtained from APEX on 3 and 6 November 2010 with a total of 352 minutes on-source\footnote{Part of ESO programs 086.A--0793 and 087.A--0815; PI: De Breuck; and 086.F--9318 and 087.F--9320; PI: Greve}.  Atmospheric conditions were excellent, with the opacity at 225 GHz in the range 0.025 - 0.041. Calibration of each spectral channel was obtained from continuum observations of Mars, Uranus, and Neptune. The spectrum is shown in Figure~\ref{fig:zspec-spectrum}, and is the source of the original redshift determination for \smg.  As a check on the calibration, we compared the SPT continuum flux at 1.4 mm with the Z-Spec flux and found them to be consistent.  

Once the redshift was established (based on the detected \coeight\ and \coseven\ lines), we performed a fit of all likely species in the Z-Spec band with the redshift held fixed at $z=2.7817$; the results are given in Table \ref{tab:lines}.  For Z-Spec, the \coseven\ line is blended with the [\ion{C}{1}](2$\rightarrow$1) due to the wide instrument channel width, and therefore the division between the line fluxes is uncertain.  The fit places approximately equal flux in the two lines: \coseven\ = $16\pm8$ Jy km s$^{-1}$ and [\ion{C}{1}](2$\to$1) = $12\pm8$ Jy km s$^{-1}$.  

In addition to the CO and [\ion{C}{1}]lines, there are a number of H$_2$O transitions which are redshifted into this band.  We identify two H$_2$O features; a $2.7\sigma$ detection of the H$_2$O(2$_{02}$$\to$$1_{11}$) line in emission (as seen in several other galaxies: \citealt{bradford09, omont11, omont13, yang13}), and a $2.7\sigma$ detection of the H$_2$O(1$_{11}$$\to$$0_{00}$) absorption line at 294 GHz, similar to the strength of that observed in Arp220 \citep{rangwala11}.

\begin{figure}
\centering 
{\includegraphics[width=9cm]{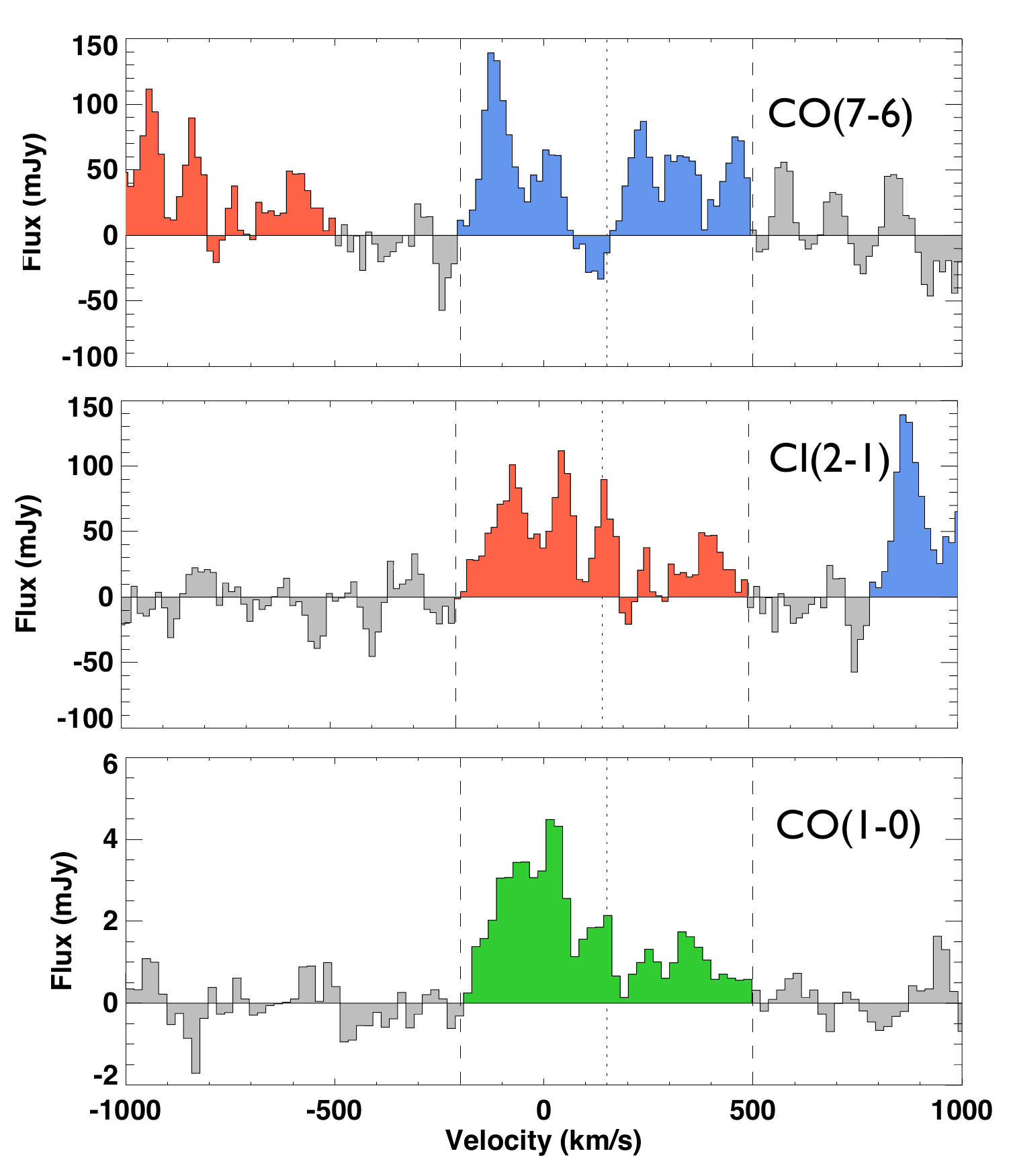}}
\caption{
Spectra of molecular emission lines in \smg, showing data from both the SMA and ATCA. All lines have been velocity-centered to a source redshift of $z=2.7817$. {\it Top panel:} The $^{12}{\rm CO}(7-6)$ emission line is detected at 213.37 GHz with the SMA, and is highlighted in blue. {\it Middle panel:} The [\ion{C}{1}]$(2-1)$ line is also detected in the same SMA track, and is shown in red. Note that this is the same SMA spectrum as shown in the top panel, but centered to a velocity appropriate for the [\ion{C}{1}]$(2-1)$ line. {\it Lower panel:} The velocity-centered ATCA 30 GHz spectrum, which detects the \coone\ line, is shown in green. All lines show a double peaked structure (indicative of distinct kinematic components), and in all lines the blue peak is stronger than the red peak. Vertical dashed lines indicate the velocity range used to integrate line fluxes, and the vertical dotted line indicates the halfway point between the two.} 
\label{smaspec}
\end{figure}

\begin{table*}
\centering
\caption{All IR emission line observations for SPT0538-50. Undetected SPIRE lines have $2\sigma$ upper limits on their luminosities quoted. All values are as observed (i.e., they have not been corrected for the magnification factor of $21\pm4$). 
Fluxes for the undetected SPIRE lines have been measured by simply integrating at the expected line position, assuming a linewidth matching the $^{12}$CO$(1-0)$ emission.}

\begin{tabular}{@{}lcccccc@{}}
\hline
Line  & $\nu_{rest}$ & $\nu_{obs}$  &   Flux& $L_{line}$ & $ L_{line}/L_{FIR}$  & Source\\
 &  GHz & GHz  & Jy km s$^{-1}$  & $[10^{9} L_{\sun}]$ &$[10^{-5}]$  & \\
 \hline
 \hline \\
\hspace{0em}[\ion{O}{3}]-52$\mu$m 	 & 5786.0  & 1530.0                 & $-25 \pm 200$    &$<70$  &$<105$  &   SPIRE-FTS\\
\hspace{0em}[\ion{N}{3}]--57$\mu$m  	 & 5230.6   & 1383.1               &  $-20\pm150$   &$<70$  &$<105$  &   SPIRE-FTS\\
\hspace{0em}[\ion{O}{1}]-63$\mu$m 	        &4744.9   & 1254.7                 & $-160\pm160$      &$<75$  &$<110$  &   SPIRE-FTS\\
\hspace{0em}[\ion{O}{3}]-88$\mu$m         & 3393.1  & 897.2                 & $150 \pm 200$    & $<50$  & $<80$ & SPIRE-FTS\\
\hspace{0em}[\ion{N}{2}]-122$\mu$m     	 & 2459.4  & 650.3                 & $30 \pm 140$      & $<40$  & $<60$   & SPIRE-FTS\\
\hspace{0em}[\ion{O}{1}]-146$\mu$m 	 & 2060.1   & 544.8                  & $-100\pm180$    &$<95$  &$<145$  &   SPIRE-FTS\\
\hspace{0em}[\ion{C}{2}]-158$\mu$m        & 1900.6  & 502.6  &$310 \pm 190$  & $<120$ & $<185$&  SPIRE-FTS \\ \\
\hline \\
$^{12}$CO(10$\to$9)  & 1151.9 & 304.6   & -- & -- & -- & Z-Spec \\
$^{12}$CO(9$\rightarrow$8)  & 1036.9 & 274.2   & -- & -- & -- & Z-Spec \\
$^{12}$CO(8$\rightarrow$7)  & 921.8 & 243.7   & $23\pm10$ & $3.1 \pm 1.4$& $4.8 \pm 2.1$  & Z-Spec \\
$^{12}$CO(7$\rightarrow$6)  & 806.7 & 213.3   & $16\pm8$ & $1.9 \pm  0.9$ & $2.9 \pm 1.5$ & Z-Spec \\
$^{12}$CO(7$\rightarrow$6)  & 806.7 & 213.3   & $26.6 \pm 4.5$  & $3.5 \pm 0.6$ & $5.4 \pm 1.0$ & SMA \\ 
$^{12}$CO(1$\to$0)  & 115.3 & 30.4  & $1.2 \pm 0.2$&$0.02\pm 0.003$ & $0.03 \pm 0.006$ & ATCA \\ \\

\hspace{0em}[\ion{C}{1}]($^3 \mathrm{P}_2$$\rightarrow$$^3\mathrm{P}_1$)  & 809.3 & 214.0   & $12\pm8$  & $1.4 \pm 0.9$ & $2.2 \pm 1.5$ & Z-Spec \\
\hspace{0em}[\ion{C}{1}]($^3 \mathrm{P}_2$$\rightarrow$$^3\mathrm{P}_1$)  & 809.3 & 214.0  &$19.7\pm2.6$   & $2.3 \pm  0.3$ & $3.5 \pm 0.6 $& SMA \\ \\

H$_2$O(3$_{12}$$\rightarrow$$2_{21}$)  & 1153.1 & 304.9   & -- & -- &  -- & Z-Spec \\
H$_2$O(1$_{11}$$\rightarrow$$0_{00}$)  & 1113.3 & 294.4   &  $-19\pm7$ & --&  -- & Z-Spec \\
H$_2$O(3$_{12}$$\rightarrow$$3_{03}$)  & 1097.4 & 290.2   & -- & --&  -- & Z-Spec \\
H$_2$O(2$_{02}$$\rightarrow$$1_{11}$)  & 987.9 & 261.2  & $19\pm7$&  $ 2.8 \pm 1.0$ &$4.3 \pm 1.6$  & Z-Spec \\
H$_2$O(2$_{11}$$\rightarrow$$2_{02}$)  & 752.0 & 198.8  & -- & -- & -- & Z-Spec \\ \\

\hline \\
\end{tabular}
\label{tab:lines}
\end{table*}

\subsubsection{SMA}
\label{sec:sma}



 \smg\ was observed by the Submillimeter Array (SMA) in the compact configuration on 2010 November 9 and 27.  The total on-source integration time was two hours.  The upper and lower sidebands were centered around 225 and 213GHz, respectively.  Flux calibration was obtained by observing Callisto (November 9) and Uranus (November 27), while the quasar 3C84 was used for passband calibration.  Complex gain calibration and  
astrometric verification were provided by the quasars J0538-440 and J0526-4830, respectively.  Baselines  ranged from 5--55k$\lambda$, and the synthesized beam FWHM was $5''\times17''$.  The redshifted \coseven\ and [\ion{C}{1}]($^3P_2 - ^3 P_1$) lines (hereafter [\ion{C}{1}]$(2-1)$) were located in the lower sideband, while continuum information was derived from the upper sideband and line-free channels of the lower sideband.  The correlator was configured for 3.25MHz channels and 4GHz total bandwidth, corresponding to 4.6 km s$^{-1}$ spectral resolution over the 5300 km s$^{-1}$ band.

The SMA data were reduced using the MIRIAD software package \citep{sault95}. 
Continuum images were created using multi-frequency synthesis to combine both sidebands, then CLEANed using natural weighting to 2$\times$ the noise level to remove sidelobe effects.  Spectral datacubes were produced by averaging velocity channels to 90 km/s resolution before inverting to the image plane.  

The SMA 220GHz spectrum is shown in Fig. \ref{smaspec}. The \coseven\ and [\ion{C}{1}]$(2-1)$ lines are each detected at $\sim 6\sigma$. We have shown a velocity-zeroed overlay of both detected emission lines (where zero velocity is defined at the targeted redshift, $z=2.7817$, as discussed in \S\ref{sec:redshifts} below).


\subsubsection{Herschel SPIRE FTS}
\label{sec:spire}

\begin{figure*}
\centering
{\includegraphics[width=17cm]{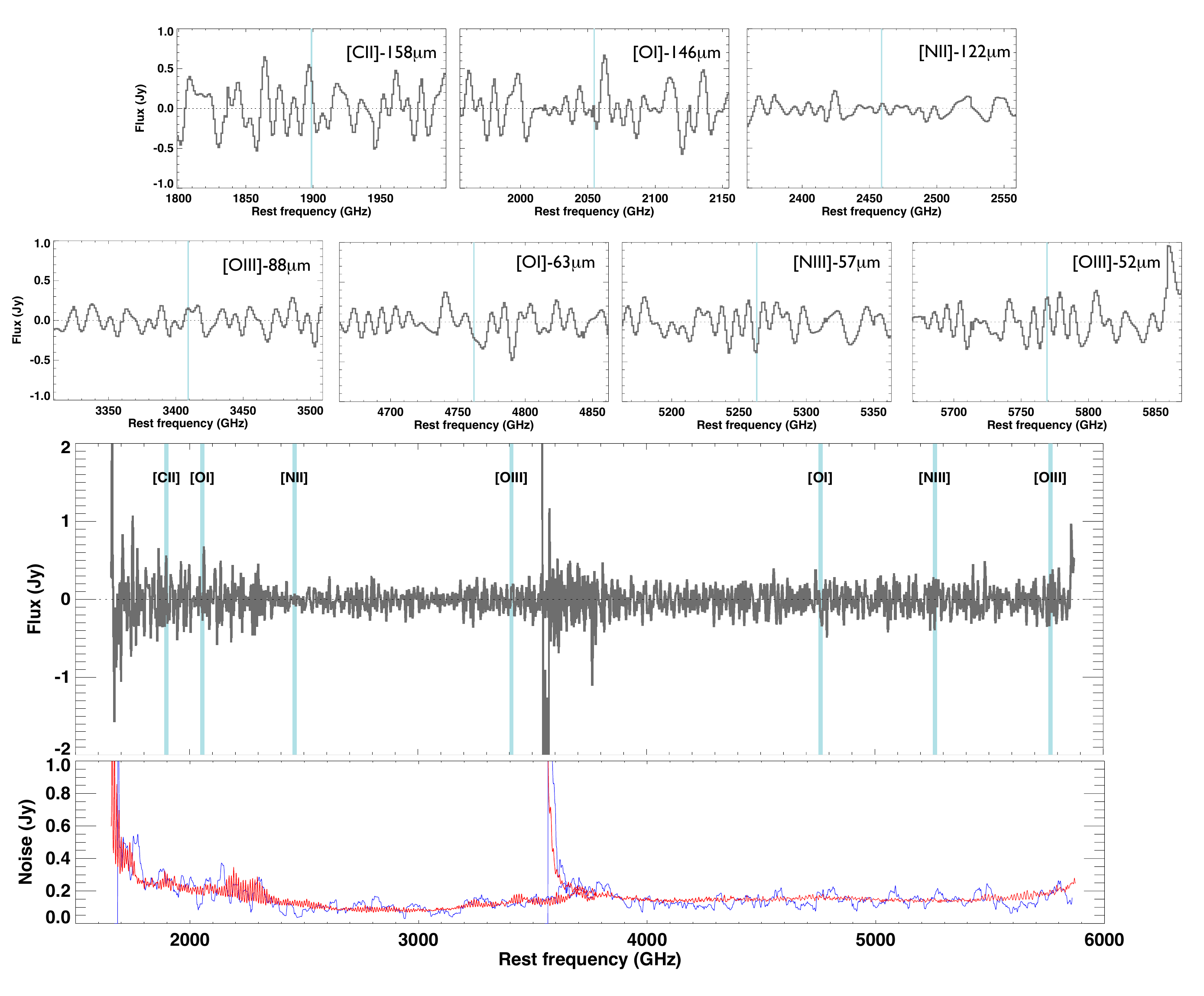}}
\caption{{\it Herschel} SPIRE FTS spectrum of \smg. 
The center panel shows the spectrum of \smg. The expected positions of some spectral lines are indicated. The upper panels show cutouts at the expected position of the lines, zoomed to show $\pm$ 100 GHz around the line frequency. The lower panel shows the noise spectrum, both the uncertainty on the mean of the repeated scans (red), and the dispersion within $\pm 5000$ km s$^{-1}$ of each channel (blue).}
\label{fig:spirespec}
\end{figure*}

\smg\ was observed with the SPIRE Fourier Transform Spectrometer on 2011 November 8, in single pointing mode as part of program OT1\_dmarrone\_1. Both observing bands were used; the short wavelength band, covering $194 - 313 \mu$m (1544 - 959 GHz),
 and the long wavelength band covering $303 - 671 \mu$m (989 - 467 GHz). 
Observations were made in the high spectral resolution mode, resulting in a resolution of 0.04 cm$^{-1}$ (1.2~GHz) over each of the two bands. One hundred repetitions were carried out, with a resultant on-source integration time of 13752 s. 

The data were processed and calibrated using the standard SPIRE FTS pipeline. \smg\ is compact relative to the SPIRE FTS beam -- our ALMA observations suggest the extent of the source is, at most, $5''$ in extent in the far-IR, while the effective SPIRE beam varies from $\sim17''$ to $\sim42''$ across the spectrum. As such, we used calibration scripts appropriate for the reduction of a pure point source. For this relatively faint galaxy, the measured spectrum is dominated by emission from the instrument itself. This was removed using a combination of techniques, including subtraction of the average of adjacent spectral pixels and subtraction of a `dark sky' reference spectrum with the same integration time. 

The SPIRE FTS spectrum of \smg\ is shown in Fig. \ref{fig:spirespec}. A spectral baseline has been removed using a boxcar average over 30 channels. We fail to robustly detect any of the ISM lines which fall into the observed spectral range. We derive $2\sigma$ upper limits on line fluxes and luminosities following \cite{valtchanov11}, by taking the expected line peak to be 2 times the RMS noise (calculated within $\pm 5000$ km s$^{-1}$ of the expected line centroid). 
Table \ref{tab:lines} gives upper limits for the six ISM fine structure lines falling within the spectral range of our observations. We also give the value of the velocity-integrated flux at the position of each of the far-IR fine structure lines (where the velocity range has been chosen following the width of the \coone\ line). 

There is a $\sim 2\sigma$ (relative to the local rms) excess of flux at the expected position of the [\ion{C}{2}]-158$\mu$m line, which can also been seen in a visual examination of the spectrum (see zoomed cutouts, Fig. \ref{fig:spirespec}). Integrating the spectrum at the expected position of the [\ion{C}{2}]-158$\mu$m line, assuming a line-width matching the \coone\ emission, results in a flux density of $310 \pm 190$ Jy km s$^{-1}$. Here, we treat [\ion{C}{2}]-158$\mu$m as a non-detection, and simply note the possible presence of a weak emission feature.



\subsubsection{VLT XSHOOTER}
\label{sec:xshooter}

 

\begin{figure*}
\centering
{\includegraphics[width=15cm]{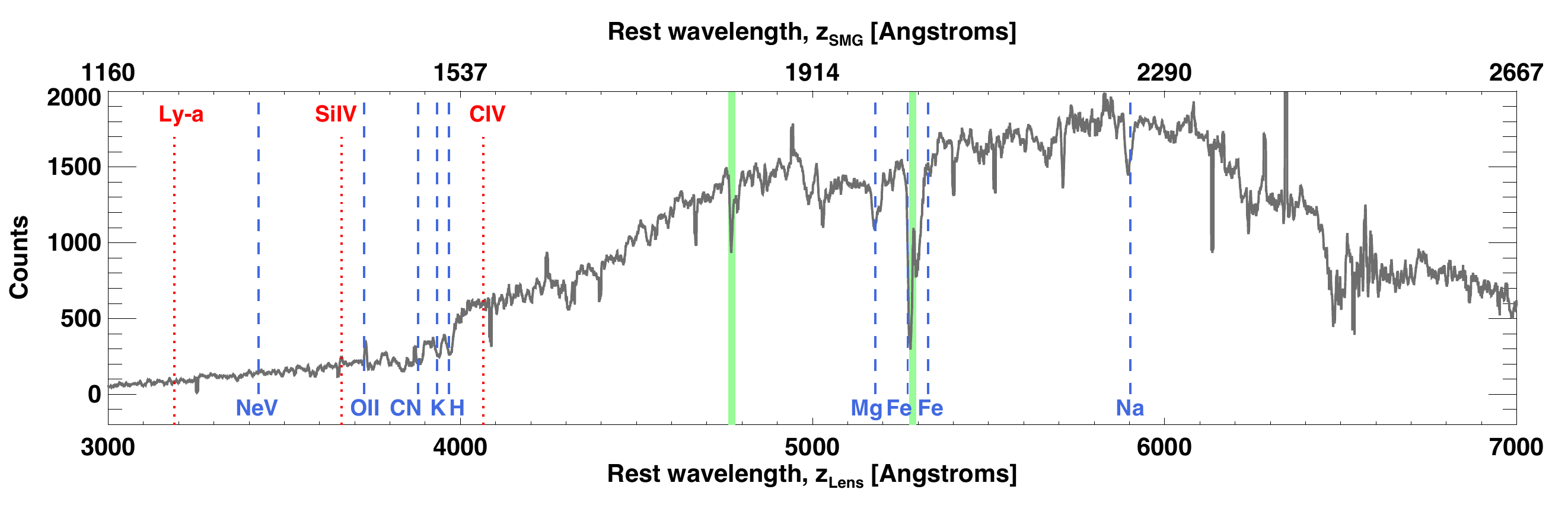}}
\caption{VLT/XShooter spectra of \smg. The lower axis shows the rest wavelength at the redshift of the foreground lens ($z=0.441$), while the upper axis shows the rest wavelength at the redshift of \smg\ ($z=2.7817$). Spectral lines marked in blue are shown in the rest frame of the foreground lens, while lines marked in red are shown in the rest frame of the DSFG. The atmospheric A and B absorption bands are marked in green. }
\label{fig:xshooter}
\end{figure*}

\smg\ was observed by {\it VLT}/XSHOOTER on 2010 March 16 (PID 284.A-5029), for a total of $6 \times 900$ sec integrations (see \citealt{vernet11} for instrument details). Half of these were taken in poor weather, however, and the spectrum presented here uses only 3 such integrations. The slit widths used were $1''$, $0.9''$, and $0.9''$ in the UV, Visible, and NIR respectively.
Further details of the observations are discussed in \cite{hezaveh13}. Fig. \ref{fig:xshooter} shows the composite XSHOOTER spectrum of \smg, with the wavelengths of lines (detected or undetected) in the source and lens labelled. \\ \\ \\

\section{Analysis}
\label{sec:results}

\subsection{Redshift}
\label{sec:redshifts}

The redshift of \smg\ is not usefully constrained by the millimeter-wavelength photometry from SPT alone. An initial redshift search was carried out using Z-Spec, using the technique described in \cite{lupu12} (which deals with low signal to noise lines by combining the significance of several detected lines).  Several ISM emission lines were detected, including two high-$J$ transitions of $^{12}$CO: \coseven\ and \coeight. Based on these lines, the redshift was determined to be $z=2.783$. 

This initial redshift was subsequently independently confirmed by the detection of the \coseven\ and [\ion{C}{1}] lines by the SMA, and the strong detection of the CO(1$\to$0) line with ATCA (as discussed in \citealt{aravena13}), which both yielded a  redshift of $z=2.7817$.

We use optical spectroscopy to determine the redshift of the foreground lens, as reported in \citealt{hezaveh13}. The {\it VLT}/X-SHOOTER spectrum, shown in Fig. \ref{fig:xshooter}, contains a multitude of absorption lines due to the foreground galaxy, and a redshift of  $z=0.441$ was determined by fitting Gaussian profiles to the Ca H+K absorption lines.

In addition to the foreground absorption lines, the ionized silicon line (Si {\sc iv} at 1397\AA) can be identified as originating from \smg, complete with a well-resolved P Cygni profile (characteristic of a rapid outflow). This identification would imply a redshift offset; the silicon line is detected bluewards of the wavelength expected for $z$=2.7817. The peak (i.e., of the red side) of the Si {\sc iv} P Cygni profile is detected at 5427\AA, implying a line redshift $z=2.7692$. If interpreted as an outflow, this redshift offset corresponds to a velocity shift of $ v \sim 1000$ km s$^{-1}$.

This velocity offset can be compared to values in the literature for similar galaxies. \citet{banerji11} presents a correlation between outflow velocity and star formation rate for a variety of galaxy types. From their compilation, it can be seen that a star formation-driven outflow of $\sim 1000$ km s$^{-1}$ is certainly not unusual for a galaxy such as \smg\ with a star formation rate of 760 M$_{\sun}$/yr (see \S \ref{SED} below). 

Other than Si {\sc iv}, however, we do not detect any optical emission lines (such as Lyman-$\alpha$ or C {\sc iv} 1550\AA ) from \smg. The non-detection of Lyman-$\alpha$ is unsurprising, due to the likely high column density of gas and dust present in a starbursting system such as \smg.  

\subsection{NIR morphology, and source/lens de-blending}

A major goal of our imaging program was to examine the morphological structure of \smg. However, while observations at millimeter and submillimeter wavelengths provide an essentially unobscured view of the background DSFG, in the NIR (i.e., {\it HST}/WFC3 and {\it Spitzer}/IRAC) any emission from the background DSFG is strongly blended with that from the foreground lens. If we are to uncover the structure of the DSFG it is therefore necessary to de-blend the image, subtracting emission from the lens galaxy and leaving only the DSFG.

\subsubsection{Modelling the foreground lens}

We first create a model of the lens as observed by {\it HST}/WFC3 (using a co-add between the F110W and F160W bands). For this purpose, we chose the fitting code GALFIT 3 \citep{peng10}, which can 
simultaneously fit several source components and multiple neighboring galaxies (see \citealt{haussler07}) -- vital for modeling a galaxy in a crowded field such as \smg. If neighboring galaxies are not simultaneously modeled, the resulting fits can be poor. 

We simultaneously fit the central elliptical galaxy and the nine closest neighbors. We attempted to fit the central galaxy with a standard 2-component model, consisting of a central bulge plus an exponential disc. Fits with this two-component model were generally poor, with unphysical parameters required in the model (convergence was only achievable with a bulge S\'ersic index $n_b>6$), and significant central structure remaining in the residuals. 
A three-component model, including a central point source, provided a superior fit, with physically-plausible parameters for each component and little residual structure. The effective radius of the `point source' component fitted by GALFIT is $750\pm 20$ pc. We allowed the S\'ersic index of the bulge to vary as a free parameter, with the result that the best-fitting model had a bulge S\'ersic index of $n_b=2.65$.

The top panels of Fig. \ref{fig:ir} show the original {\it HST}/WFC3 image (top left), the final `best-fit' model of the WFC3 data (top middle), and the residuals after subtracting this best-fit model from the WFC3 data (top right). We see no trace of the DSFG (the structure seen by ALMA) in these residuals, from which we conclude that the source detected in our {\it HST}/WFC3 imaging consists of flux from the low-$z$ lens galaxy alone. 

We note that the source as detected in IRAC (both 3.6$\mu$m and 4.5$\mu$m) emission is extended on scales larger than those seen in the WFC3 image. The effective (i.e., half light) radius as measured in the IRAC images is 2.0$''$ (= 11.3 kpc at $z=0.441$). In contract, in each of the WFC3 images the effective radius is 1.4$''$ (= 7.9 kpc at $z=0.441$). As the WFC3 residuals show no trace of the background SMG, this size difference further suggests the existence of blended emission from a background galaxy being detected in our IRAC images.

In addition, we also see no trace of emission from \smg\ remaining in the VLT/ISAAC Ks-band data. While this data has lower angular resolution than the HST images (and so is less suited to modelling the lens galaxy), we can use this Ks-band non-detection to place an upper limit on the flux from \smg: based on the rms noise in the ISAAC map, we place a Ks-band upper limit on \smg\ of $<20.19$ mag.

\subsubsection{De-blending the source/lens emission}

\begin{figure*}[!hbp]
\centering
{\includegraphics[width=17cm]{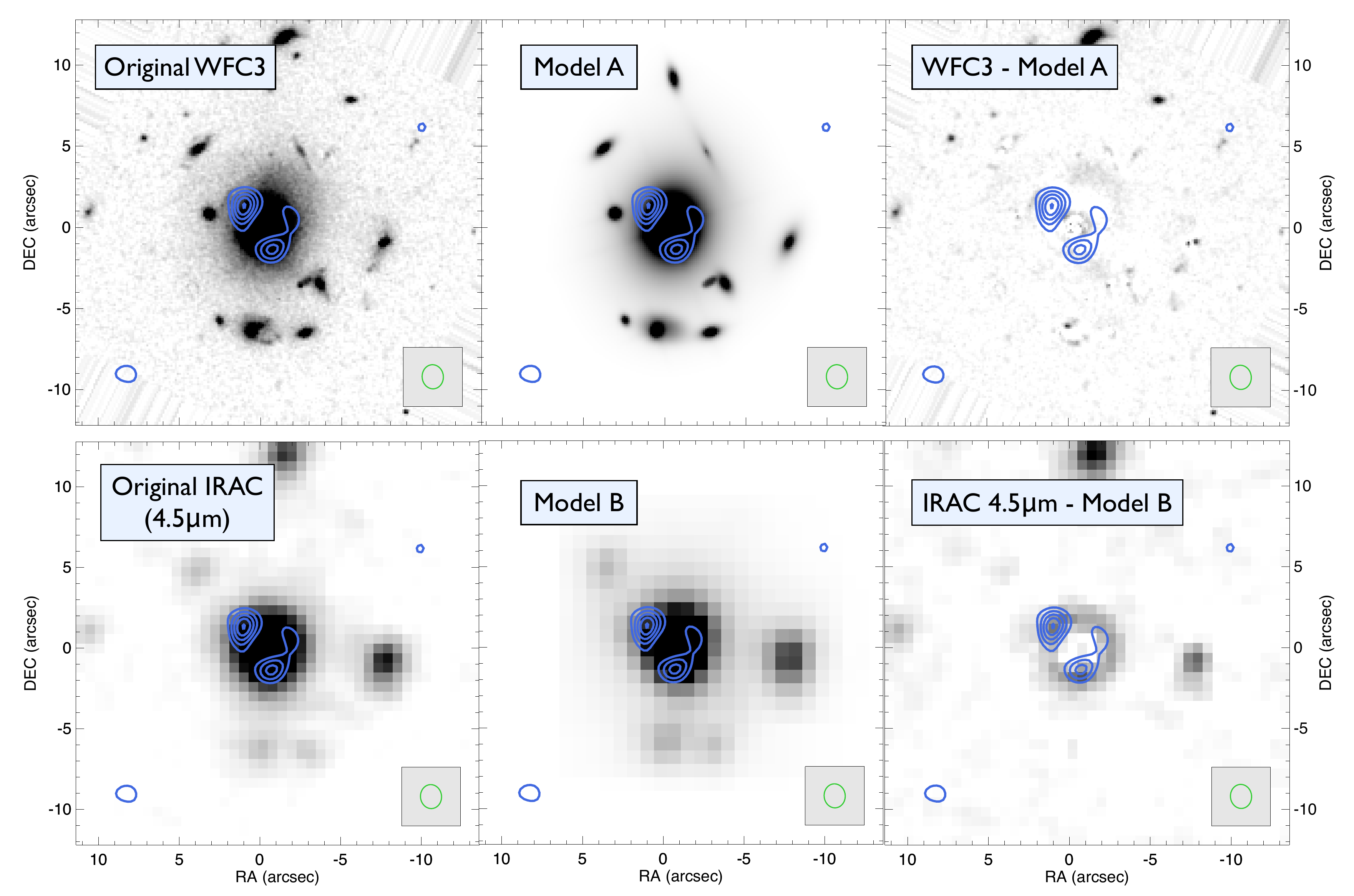}}
\caption{IR imaging of \smg, showing our de-blending technique. In each $25\arcsec\times25\arcsec$ panel, blue contours show sub-mm continuum emission imaged by ALMA (see \S \ref{sec:alma}). Contours are in steps of $2\sigma$, starting at $3\sigma$.  {\it Top Left:} {\it HST}/WFC3 image of \smg, created by co-adding images from the F110W and F160W filters. The high-$z$ DSFG is totally hidden by the lens galaxy in the foreground. {\it Top Center:} The best-fitting GALFIT model of \smg\ (`model A'). {\it Top Right:} {\it HST}/WFC3 image of \smg\, after the subtraction of the best GALFIT model. \\
The lower row shows our de-blending technique.
{\it Lower Left:} {\it Spitzer}/IRAC $4.5 \mu$m image of \smg. {\it Lower center:} The GALFIT model, convolved with the IRAC PSF at $4.5 \mu$m (`model B').  Each object has been allowed to have a different HST-IRAC color. {\it Lower Right:} {\it Spitzer}/IRAC $4.5 \mu$m image of \smg, after subtraction of the convolved GALFIT model. 
The Einstein ring structure of the DSFG, as traced by submillimeter continuum emission, is clearly visible in both IRAC channels once the foreground galaxy is removed. 
}
\label{fig:ir}
\end{figure*}

The next step is to construct a model of the lens galaxy as seen by IRAC. We take the high-resolution model of the lens galaxy developed from the {\it HST}/WFC3 imaging, and convolve with the IRAC point-spread function (PSF). For this, we used an empirically-derived PSF, for which we used a uniformly-weighted stack of ten bright, isolated stars in the same field. This model IRAC image can then be subtracted from the true IRAC image, leaving -- as residual emission -- any flux from the DSFG which was originally blended with the lens. The lower panels of Fig. \ref{fig:ir} show these steps -- the original IRAC image (lower left), the above model convolved with the IRAC PSF (lower middle), and the residuals after subtracting this convolved model from the IRAC data (lower right).

It is of critical importance to determine the correct 3.6/4.5~\um\ normalization of the convolved galaxy model in order to recover lensed emission in the IRAC image. The normalization cannot be determined by making a direct comparison to the IR radial profile, or by the standard approach of minimizing the image residual, because the relatively large IRAC PSF will spread the contribution of any IR emission from \smg\ across much of the face of the lensing galaxy. To include the DSFG contribution to the image, we add a ring component at the radial distance of the submillimeter components identified in the ALMA image, 1.8\arcsec. It is important to note that this ring is a ``dummy'' component, and serves only as a guide to finding the optimum normalization of the model lens galaxy; it is not used in the final image subtraction, and so it does not contribute to any potential residuals. 

The morphology of the DSFG is recovered once the properly normalized lens model image is removed.
We varied the normalization of each component separately, using simultaneous $\chi^2$ fitting to find the optimum values, determined by comparing to the true IRAC image.
\footnote{This differs somewhat from the approach of \citet{hopwood11}, who used the peaks in their submillimeter images to locate S\'ersic-profile DSFG components in the IR model. Here we make a more conservative assumption, namely that any lensed IR emission will appear somewhere on the ring that threads the submillimeter components, but not necessarily at the same positions.  As such, our dummy ring component has equal flux at all points.} 

When convergence was reached, we subtracted the resulting best fitting model IRAC image from the real IRAC data. 
Having removed flux resulting from the foreground lens, we attribute the residuals left after this subtraction to the DSFG itself. We stress that ring-shaped residuals spatially matching the ALMA submm emission are not an artefact, and are recovered even when using cruder normalization methods that do not use the dummy ``ring'' component.  (See Appendix B for discussion and figures).




The results of this image deconvolution process are shown in Fig. \ref{fig:ir}. ALMA submillimeter continuum contours have been overlaid on all images, to indicate the position and structure of the lensed DSFG emission. The close spatial match between the ALMA continuum emission and the post-subtraction IRAC residuals strongly suggests that this residual NIR emission does indeed originate from the background DSFG. 

We note that the poor subtraction of the galaxy $\sim 5\arcsec$ to the West of \smg\ is likely due to it being badly fit by a single S\'ersic-profile -- as it is significantly separated from \smg, however, it will not affect our measured residual flux and we do not attempt to remove it more thoroughly. 

We perform aperture photometry on the $3.6\mu$m and $4.5\mu$m residual images, extracting the flux within a circular aperture of radius 4$''$, and derive resultant flux densities of $S_{3.6} = 22 \pm 5 \; \mu$Jy, and $S_{4.5} = 47 \pm 8 \; \mu$Jy. Uncertainties here include both the standard photometric uncertainty and the uncertainty resulting from the deconvolution process, as described above. In \S4.3 below we present these IRAC flux densities, de-magnified using our lens model.

The [3.6]$\mu$m $-$ [4.5]$\mu$m color also provides evidence that the residual ring structure is a high-$z$ background source. The lens galaxy has a [3.6] $-$ [4.5] color of $0.43 \pm 0.04$ -- fairly typical for an early-type galaxy at intermediate redshift. The residual structure, which we attribute to the background DSFG, has a far redder [3.6] $-$ [4.5] color of $1.27 \pm 0.5$, again suggesting that the residuals are not part of the lens galaxy, but have a separate, high-$z$, origin. A [3.6] $-$ [4.5] color this red may be indicative of the presence of an AGN in \smg\ (see \citealt{yun08}), though with such large uncertainties this conclusion remains speculative.


\begin{figure}[!hbp]
\centering
{\includegraphics[width=8cm]{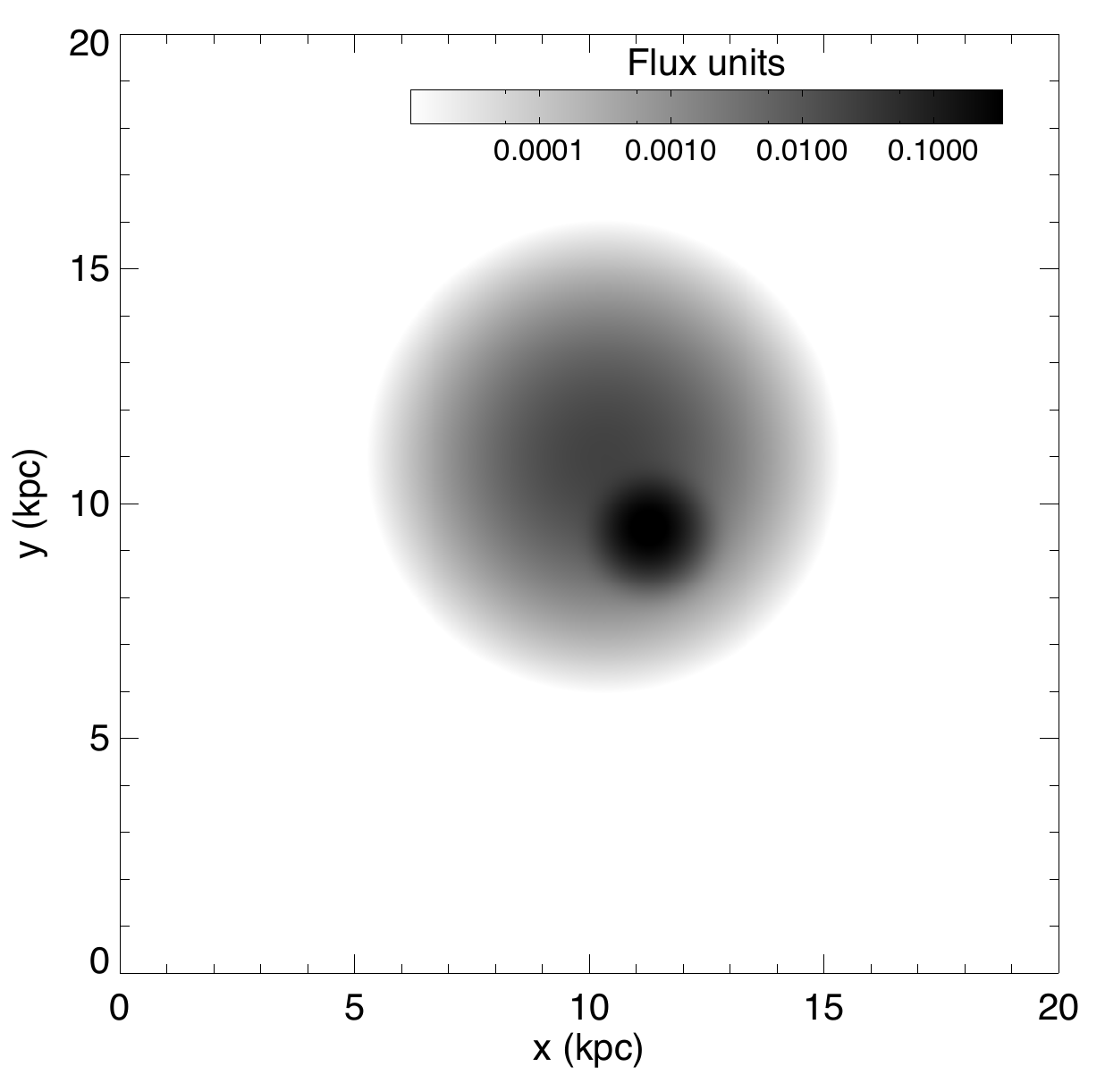}}
\caption{Source-plane reconstruction of \smg\  at 850$\mu$m, based on the lens modeling reported in \cite{hezaveh13} and described in \S\ref{sec:lensmodel}. Single-component fits to the lensed data were poor, with significant ($\sim 6 \sigma$) residuals to the South East. This two-component model provided a better fit to the data. The source consists of a large, diffuse component (responsible for 30\% of the total luminosity), and a compact, luminous component (responsible for the remaining 70\%). }
\label{fig:model}
\end{figure}

\subsection{Lens Model}
\label{sec:lensmodel}

\smg\ was modeled in detail by \cite{hezaveh13}, based on our Cycle 0 ALMA Band 7 imaging. Details of the modeling technique can be found in that work. To summarize, the modeling is carried out in the $u,v$ plane by modeling the source as a 2-D Gaussian intensity profile, predicting $u,v$ visibilities, and minimizing a $\chi^2$ figure of merit. 



As discussed by \cite{hezaveh12b}, the large size of the Einstein ring resolved by ALMA (diameter $\sim 1.8''$; see Fig.\ \ref{fig:ALMA}) suggests a particularly massive lensing halo, with a mass of $(7.15 \pm 0.05) \times 10^{11}$ M$_{\sun}$ enclosed within the Einstein radius. This massive foreground galaxy, combined with an optimal source/caustic configuration, results in a large magnification factor, $\mu = 21 \pm 4$. 
Fits using just a single component model in the source plane resulted in a significant ($>6 \sigma$) residual structure to the South-East of the image, evidence that a two-component model is required to reproduce the ALMA data. The best fitting model of \smg\ consists of two bodies, one compact ($R_{1/2} = 0.52 \pm 0.12$ kpc) source which is responsible for $70$\% of the total source-plane luminosity, and one more extended ($R_{1/2} = 1.61 \pm 0.33$ kpc) source which is responsible for the remaining $30$\%. Fig. \ref{fig:model} shows the source-plane reconstruction of \smg .

Using the lens model, it is now possible to begin reconstructing the intrinsic properties of \smg. It can be seen that de-lensing its prodigious observed submillimeter flux density, $125 \pm 7$ mJy at 870$\mu$m, results in a flux density of $6.1 \pm 1.2$ mJy\footnote{Uncertainties on this value, and on all de-lensed quantities quoted hereafter, include the uncertainty contribution from the lensing model.}, relatively modest for a submillimeter galaxy. It is worth noting that the intrinsic submillimeter flux density still falls above the formal selection limit for the original SCUBA survey \citep{smail97}, and so this source would have been identified as a `classical' DSFG even if unlensed. \\ \\

\subsection{SED fitting to IR data}
\label{SED}

\begin{figure}
\centering
{\includegraphics[width=8cm]{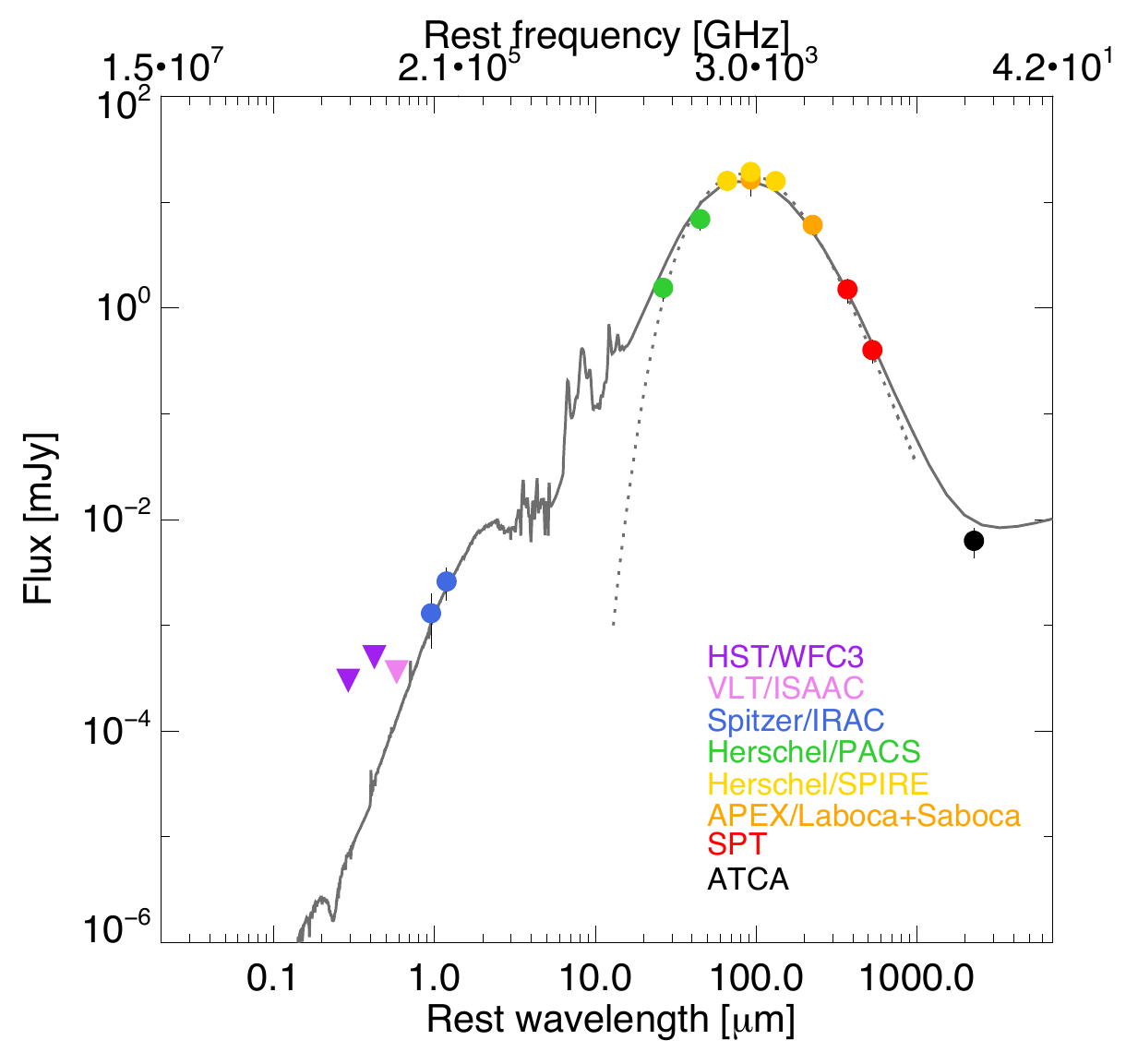}}
\caption{The result of SED fitting our \smg\ photometry, using the SED fitting code CIGALE. Fluxes are innate (i.e., demagnified using our lens model). Datapoints are ATCA 35 GHz continuum, SPT 2.0+1.4mm, APEX 350$\mu$m+870$\mu$m, {\it Herschel}/SPIRE 250$\mu$m+350$\mu$m+500$\mu$m, {\it Herschel}/PACS 100$\mu$m+160$\mu$m,  {\it Spitzer}/IRAC 3.5$\mu$m+4.5 $\mu$m and 2$\sigma$ upper limits from HST/WCF3 and VLT/ISAAC. The dashed line shows a greybody fit to the far-IR data with a dust temperature of $39 \pm 2$ K (and $\beta =2$).}
\label{fig:SED}
\end{figure}

We use the galaxy SED fitting code CIGALE (\citealt{burgarella05}; \citealt{noll09}) to fit the near- and far-IR photometric data. Table \ref{tab:IRphot} gives de-magnified fluxes for all our observed IR datapoints, which sample both the stellar mass and the dust component of \smg. We do not fit to the low frequency ($<10 $ GHz) ATCA data (given in \citealt{aravena13}), as CIGALE does not extend its fitting beyond the far-IR. 

The best-fitting SED, along with the de-magnified IR datapoints, is shown in Fig. \ref{fig:SED}. The fit was achieved assuming a solar metallicity, an old stellar population with a mean age of 2 Gyr, and a young stellar population with a mean age of 20 Myr (these three input parameters are fixed). We also assume that the older stellar population has an exponentially declining star formation history (SFH), while the young stellar population has a `box'-like SFH. 

From these fixed input parameters, we then used CIGALE to find the best-fitting star formation rate, stellar mass, and IR luminosity.
Being constrained by the FIR region of the SED, the star formation rate of \smg\ remains approximately constant when varying our input parameters. The stellar region (in the rest-frame optical/UV) is less well constrained, with only the two IRAC data points available. As a result, the derived stellar mass is more sensitive to the choice of input parameters. The stellar mass is only weakly dependent on the SFH assumed, but is more dependent on the assumed age for the young stellar population. SED fits with ages $<15$ Myr were unable to match the derived IRAC fluxes. Varying the adopted young stellar age between $15-50$ Myr causes the resulting stellar mass to vary by $\sim 50\%$. The stellar mass is also dependent on the assumed `burst fraction' -- the mass fraction of the young stellar population. We assumed a fixed burst fraction of 0.3 (as with the stellar ages above, burst fractions of $<0.3$ provided unsatisfactory fits). It is difficult to constrain the burst fraction without more extensive optical photometry; however, allowing the burst fraction to vary within a reasonable range (0.3-0.6) alters the resulting stellar mass by less than 25\%.


The SED suggests an IR luminosity (from $8-1000 \mu$m) for \smg\ of $(3.4 \pm 0.3) \times 10^{12}$ L$_{\sun}$, and an associated star formation rate of $760 \pm 100$ M$_{\sun}$/yr. These values are in line with those calculated by fitting a simple greybody to the far-IR points alone; once corrected for the lensing factor used in this work, the IR luminosity of \smg\ derived by \cite{greve12} is $(3.17 \pm 1.0) \times 10^{12}$ L$_{\sun}$. The SED fit also suggests a stellar mass of $(3.3 \pm 1.5) \times10^{10} \; {\rm M}_{\sun}$. We note that this is roughly consistent with the value of $(4.6 \pm 2.1) \times10^{10} \; {\rm M}_{\sun}$ reported in \cite{aravena13} which was estimated using an empirical relation between specific star formation rate (sSFR = SFR/M*) and radio spectral flattening as observed for local compact starbursts (see \citealt{murphy13}). From these results, we also derive a sSFR for \smg\ of $25\pm 12$ Gyr$^{-1}$. \\ \\ \\

\begin{table*}
\centering
\caption{IR and millimeter fluxes for \smg, both observed and de-magnified by the lensing magnification of $21 \pm 4$. Quoted uncertainties include photometric flux uncertainties, magnification uncertainty (where applicable), and absolute flux calibration uncertainties.}

\begin{tabular}{@{}ccccc@{}}
\hline
Wavelength & Observed Flux density & Innate Flux density & Source &Reference\\
 (observed) $ \lambda $ &  [mJy] & [mJy]& \\
 \hline
 \hline
 1.1$\mu$m	& $<2.4 \times 10^{-4} $ & $< 5.0 \times 10^{-3}$  &{\it HST}/WFC3 & This work\\
 1.6$\mu$m	& $<4.0 \times 10^{-4} $ & $< 8.4 \times 10^{-3}$  &{\it HST}/WFC3 & This work\\
  2.2$\mu$m	& $<2.6 \times 10^{-4} $ & $< 5.5 \times 10^{-3}$  &{\it VLT}/ISAAC & This work\\

 3.6$\mu$m	& $0.022 \pm 0.005$ & $0.0011 \pm 0.0005$  &{\it Spitzer}/IRAC & This work\\
4.5$\mu$m	& $0.047 \pm 0.008$ &$0.0023 \pm 0.0006$   &{\it Spitzer}/IRAC & This work\\
100$\mu$m	& $31 \pm 2$  & $1.5 \pm 0.3$  &{\it Herschel}/PACS & This work\\
160$\mu$m	& $141 \pm 15$&$6.7 \pm 1.5$   &{\it Herschel}/PACS & This work\\
250$\mu$m	&$326 \pm 23$ & $15.5 \pm 3.1$  &{\it Herschel}/SPIRE & This work\\
350$\mu$m	& $396 \pm 38$&$18.9 \pm 3.9$   &{\it Herschel}/SPIRE & This work\\
500$\mu$m	& $325 \pm 24$& $15.5 \pm 3.2$  &{\it Herschel}/SPIRE & This work \\
350$\mu$m	&$336 \pm 88$ &$16.4 \pm 5.1$   &APEX/SABOCA & \cite{greve12} \\
870$\mu$m 	& $125 \pm 7$&$6.1 \pm 1.2$ & APEX/LABOCA & \cite{greve12}\\
1.4mm          & $28 \pm 4.6$ &$1.5 \pm 0.4$&SPT & \cite{vieira10} \\
2mm             & $8.5 \pm 1.4$ &$0.4 \pm 0.1$  & SPT & \cite{vieira10}\\
8.6mm          & $0.13 \pm 0.02$ & $0.006 \pm 0.002$ & ATCA & \cite{aravena13} \\
\hline \\
\end{tabular}
\label{tab:IRphot}
\end{table*}

\section{Physical Properties}
\label{sec:properties}

\subsection{Source Structure}

As described above, \smg\ is best modeled as being comprised of two source-plane components: a luminous compact body with $R_{1/2} = 0.52 \pm 0.12$ kpc, and a more diffuse body with $R_{1/2} = 1.61 \pm 0.33$ kpc (models assuming just a single component were a poor fit to the data, resulting in $>6 \sigma$ residuals). We note that the lensing models here assume symmetric Gaussian profiles for the source components, and as such do not take into account the likely complexity of the true galaxy. 

Being a luminous DSFG, \smg\ is potentially undergoing a major merger event which is driving the high rate of star formation (e.g. \citealt{engel10}). However, it is not immediately obvious whether the two components revealed by the lensing reconstruction correspond to separate bodies involved in a merger, or whether the compact, luminous component simply represents a site of star formation embedded within a more diffuse ISM. 

Taking the two source-plane components of \smg\ separately, we use the projected radii and luminosities to calculate the star formation rate density in each of the two components of \smg. By assuming that the flux density distribution revealed in the ALMA imaging (70/30 in favor of the compact component) can be applied to the total IR luminosity, we calculate that the SFR density ($\Sigma_{SFR}$) of the compact component is over an order of magnitude higher than its diffuse companion. For the compact and extended bodies, we reach values of $\Sigma_{SFR} = 630 \pm 240$ M$_{\sun}$ yr$^{-1}$ kpc$^{-2}$ and $\Sigma_{SFR} = 31 \pm 11$ M$_{\sun}$ yr$^{-1}$ kpc$^{-2}$, respectively. 

The majority (70\%) of the total star formation in \smg, therefore, is likely occurring in a compact region which has a star formation rate surface density comparable to the densest star-forming regions contained within local merging ULIRGs  (i.e., \citealt{Kennicutt98}; \citealt{genzel10}). This is suggestive of a galaxy undergoing a major merger (in which tidal torques funnel molecular material into a dense, star forming region), rather than secular star formation (which is expected to produce a more extended, disk-like system). The fact that the majority of the star formation in \smg\ is occurring in a dense, ULIRG-like environment is supported by both the highly excited CO emission discussed in \S\ref{sec:ISM} below, and the far-IR line deficits discussed below in \S\ref{sec:lines}. 

These conclusions are further supported by the derived specific star formation rate of \smg\, of $\sim 25$ Gyr$^{-1}$ (see \S\ref{SED} above). This value is approximately a factor of five above the `main sequence' at $z=2.7$ (e.g., \citealt{elbaz11, karim11}), which again suggests the presence of a major-merger driven starburst. The double-peaked molecular lines revealed by our millimeter-wavelength spectroscopy are certainly in line with this picture. However, in the absence of high-resolution kinematic  observations it is difficult to draw firm conclusions about the source structure.

\subsection{ISM Properties}
\label{sec:ISM}

The velocity-resolved measurements of CO and {\sc Ci} (Figure~\ref{smaspec}) reveal \smg\ to be a double-peaked source. Emission from the low-$J$ CO and {\sc Ci} lines is expected to be co-spatial, as the two molecules have similar critical densities and excitation temperatures, to the extent that {\sc Ci} is also expected to be a good tracer of the total molecular gas mass \citep{papadopoulos04}.

As our SMA-detected molecular lines do not have particularly Gaussian profiles, we determine line fluxes by integrating our spectra between velocity limits ($-200$ to $+500$~km~s$^{-1}$) defined by the profile of the CO($1-0$) emission line, detected with high S/N in our ATCA spectrum (see \S\ref{sec:atca}, and Fig. \ref{smaspec} above). We find the $^{12}{\rm CO}(7-6)$ line has a flux $S_{CO(7-6)} = 26.6 \pm 4.5$ Jy km s$^{-1}$, while the [\ion{C}{1}]($^3P_2 - ^3 P_1$) has a flux $S_{[CI]} = 19.7 \pm 3.1$ Jy km s$^{-1}$.

For both the CO$(1-0)$ and {\sc Ci} lines the `red' peak is suppressed relative to the `blue' peak, though this effect is less evident in the CO$(7-6)$ line. We can integrate the peaks separately (above and below 150 km/s, the point which lies between the two peaks) to assess the strength of the red peak suppression. The CO$(1-0)$ and {\sc Ci} lines each have $\sim 80\%$ of their line flux in their blue peak, while the CO$(7-6)$ line shows far more symmetry between the peaks (with the blue peak containing $\sim 60\%$ of the total line flux). The asymmetry between the red and blue peaks seen in the CO$(1-0)$ and {\sc Ci} lines suggests that they represent distinct components of an on-going major merger, as opposed to being due to the rotation of a disc.

\subsubsection{Molecular gas properties}

As discussed above in \S2.1.2, \smg\ has been observed by ATCA in \coone\ \citep{aravena13}. They report a de-lensed\footnote{The de-lensing was done using the magnification factor of $\mu = 21\pm4$ used throughout this work.} \coone\ flux of $I_{CO(1-0)} = 1.20 \pm 0.20$ Jy km s$^{-1}$, and a resulting \coone\ luminosity of $L'_{CO(1-0)} = (2.16 \pm 0.13) \times 10^{10}$ K km s$^{-1}$ pc$^{2}$.  
Using a standard ULIRG-appropriate CO/H$_2$ conversion factor\footnote{Which is itself uncertain -- the standard value for starburst galaxies, which we adopt here, is $\alpha_{CO} = 0.8$ M$_{\sun}$ (K km s$^{-1}$ pc$^{2})^{-1}$; see \cite{solomon05,carilli13}.}, this corresponds to a molecular gas mass of $1.7 \pm 0.5 \times 10^{10}$ M$_{\sun}$ (the uncertainty on this value incorporates both the uncertainty on the lensing magnification, and an estimated 15\% calibration uncertainty). This is towards the lower end of the molecular gas mass distribution for DSFGs (\citealt{bothwell13} report a mean molecular gas mass for unlensed DSFGs of $5.3 \pm 1.0 \times 10^{10}$ M$_{\sun}$), but is not incommensurate with the derived SFR and stellar mass. We calculate the IR-to-H$_2$ ratio (defined here as SFR/M(H$_2$)) of $4.4 \pm 1.4 \times 10^{-8} \; {\rm yr}^{-1}$, and a baryonic gas fraction -- defined as M(H$_2$)/(M(*)+M(H$_2$)) -- of $f_{\rm gas} \sim 0.4$.

\cite{aravena13} used the peak separation of the \coone\ line to estimate a dynamical mass for \smg. Taking the source separation (2 kpc) and the \coone\ peak separation above, they estimate a dynamical mass for \smg\ of $\sim 1.3 \times 10^{11}$ M$_{\sun}$.  

\begin{figure}[!]
\centering
{\includegraphics[width=8cm]{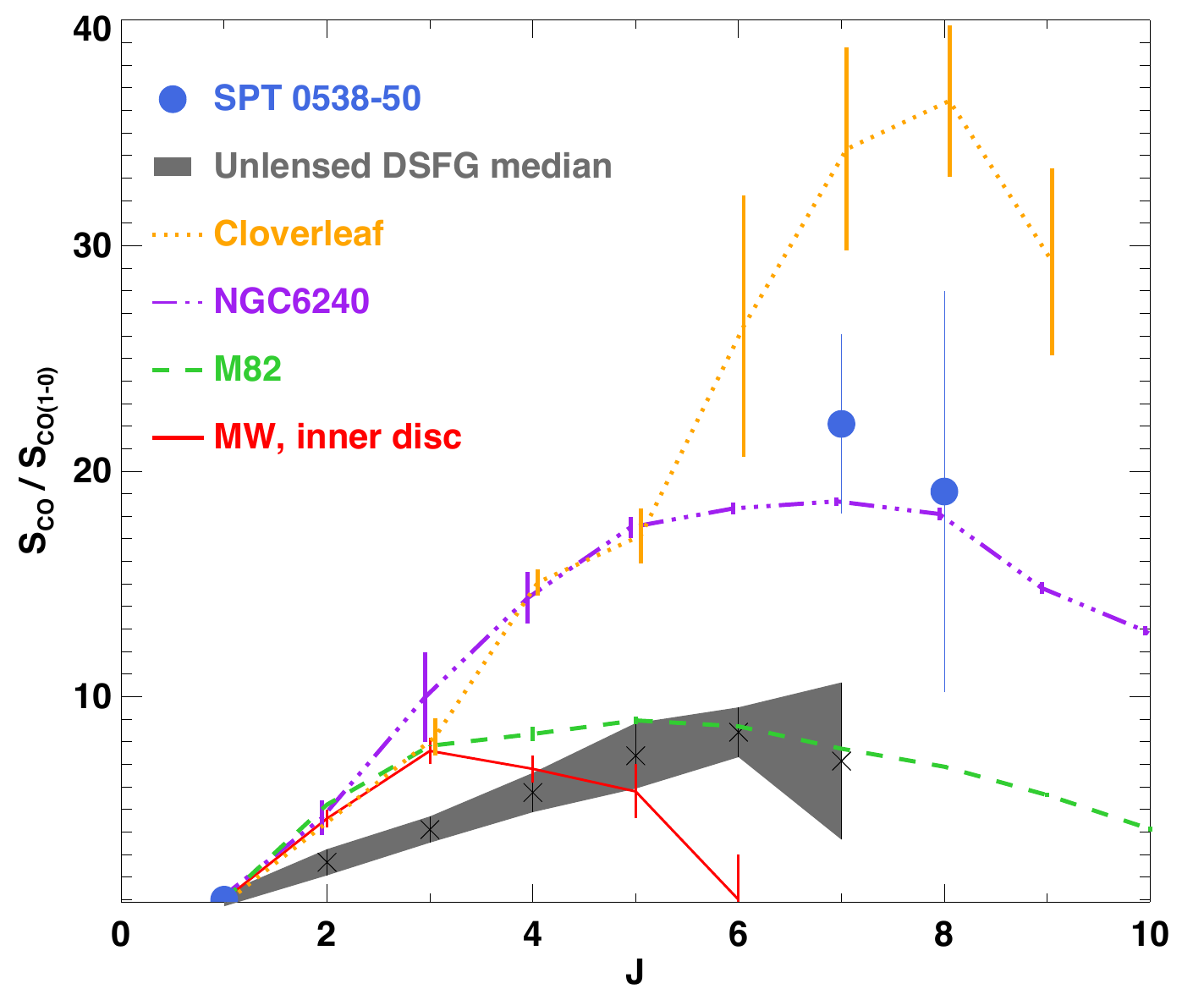}}
\caption{The CO spectral line energy distribution (SLED), normalized to the \coone\ flux, for \smg. Also shown is the median SLED for a large sample of unlensed DSFGs \citep{bothwell13}; the SLED for local LIRG NGC6240 \citep{meijerink13}, the prototypical low-$z$ starburst M82 (N. Rangwala, private communication -- see also \citealt{panuzzo10}; \citealt{kamenetzky12}), the inner disc of the Milky Way \citep{fixsen99}, and the lensed Cloverleaf QSO (\citealt{bradford09}; \citealt{riechers2011c}). \smg\ shows a high excitation towards higher $J$ relative to similar starburst galaxies. }
\label{fig:SLED}
\end{figure}

As the \coone\ line is a tracer of the cold molecular gas reservoir and the \coseven\ line traces warmer star-forming gas, we can use the ratio of the \coseven\ and \coone\ luminosities as a crude estimator of the ISM excitation in \smg. 
Figure \ref{fig:SLED} shows the CO spectral line energy distribution (SLED) of \smg, compared to SLEDs of similar objects, including the `Eyelash' DSFG as presented by \cite{danielson11}, and the median SLED for a sample of unlensed DSFGs \citep{bothwell13}. It can be seen that the CO line flux from \smg\ in our two high-$J$ lines, (8$\to$7) and (7$\to$6), is far stronger than would be extrapolated from the (1$\to$0) line alone (assuming a CO SLED comparable to other DSFGs). 
We find a ratio of  $L'_{CO(7-6)} / L'_{CO(1-0)} = 0.44 \pm 0.08$, compared to $0.12 \pm 0.005$ for the `Eyelash' \citep{danielson11}, and $0.18 \pm 0.04$ for the DSFG average \citep{bothwell13}.

In the absence of further detections of mid-$J$ lines (i.e.,  $2<J<6$) to constrain the behavior of the SLED it is difficult to draw firm conclusions about the excitation conditions of the ISM in \smg. It is possible that the strong CO emission at high-$J$ indicates a star forming ISM which is displaying a higher excitation than observed in other, star formation-dominated systems. A possible AGN could also contribute to the observed CO excitation, though this is somewhat unlikely as the bulk of the submm-selected DSFG population is likely to be starburst-dominated \citep{coppin10}.

Alternatively, the extreme excitation may also result from differential lensing of the CO transitions. As discussed by \cite{serjeant12} and \citet{hezaveh12a}, compact regions can be magnified more strongly than the diffuse ISM as a whole, by lying close to a lensing caustic. If the compact region being boosted is a star-forming `clump' (with a high star formation rate density and high excitation), the effect on the SLED of \smg\ will be to produce an apparently higher excitation, by disproportionately enhancing the contribution from a compact, bright region.

\begin{table}
\centering
\caption{Derived physical properties for \smg.}

\begin{tabular}{@{}cc@{}}
\hline
Parameter  & Value\\
 \hline
 \hline
SFR & $760 \pm 100 \; {\rm M}_{\sun}$/yr\\
M$_*$ & $(3.3 \pm 1.5) \times 10^{10} {\rm M}_{\sun}$\\
sSFR & $25 \pm 12 \; {\rm Gyr}^{-1}$ \\
M$_{\rm H2}$ & $(1.7 \pm 0.5)  \times 10^{10} {\rm M}_{\sun}$  \\
L$_{\rm IR}$ & $(3.4 \pm 0.3) \times 10^{12} \; {\rm L}_{\sun}$   \\
$\Sigma_{\rm SFR}$ (compact) & $(630 \pm 240) \; {\rm M}_{\sun}$/yr/kpc$^{2}$   \\
$\Sigma_{\rm SFR}$ (diffuse) & $(31 \pm 11) \; {\rm M}_{\sun}$/yr/kpc$^{2}$ \\
\hline
\end{tabular}
\label{tab:props}
\end{table}

\subsubsection{[CI]}

We can use our measurement of the  [\ion{C}{1}](2$\to$1) line to estimate the total mass of carbon in \smg. Following \cite{walter11}, who surveyed atomic carbon emission in 10 DSFGs, we note that the typical  [\ion{C}{1}](1$\to$0)/[\ion{C}{1}](2$\to$1) ratio (excluding sources with only upper limits on both lines) is $0.45 \pm 0.14$, corresponding to an excitation temperature of 29.1 K. Assuming this value and the observed line flux, the resulting carbon mass is $(5.6 \pm 1.7) \times 10^6$ M$_{\sun}$, comparable to measured values for other DSFGs \citep{walter11}, and carbon abundance M({\sc Ci})/M(H$_2$) = $(3.3\pm1.2)\times10^{-4}$. 
We also derive the cooling contribution of {\sc Ci} (parameterized as $L_{[CI]}/L_{FIR}$, finding a value of $(9.7 \pm 3.6) \times 10^{-6}$. This value is again similar to other DSFGs {\citep{walter11}, but higher than is observed in QSOs at similar redshifts \citep{carilli13}.  

\subsubsection{Water detection and excitation conditions}

Our Z-Spec observations also tentatively detect an emission line of water, H$_2$O(2$_{02}$$\to$$1_{11}$) at 261 GHz. With a flux of $19 \pm 7$ Jy km s$^{-1}$, this line is bright with a (de-lensed) luminosity of $1.4 \times 10^8$ L$_{\sun}$. The water emission detected from \smg\ is similar to that in other high-$z$ systems. \cite{omont13} report the detection of water in six lensed ULIRGs lying between $1.5 < z < 4$, finding that water emission lines typically exhibit flux strengths $\sim 30-50\%$ of the nearby CO emission lines. In \smg\ the H$_2$O(2$_{02}$$\to$$1_{11}$) line flux is $80\pm30\%$ of the the neighboring CO(8$\to$7) line. 

\cite{yang13} present observations of various transmissions of water for a large sample of local and high-$z$ galaxies (including lensed ULIRGs at high-$z$). All transitions of water show a good correlation with IR luminosity -- for their sample of galaxies, the H$_2$O(2$_{02}$$\to$$1_{11}$) line has a mean L(H$_2$0)/L(IR) ratio of $7.58 \times 10^{-6}$, with low scatter. The water emission from \smg\ is comparatively over-luminous: the L(H$_2$0)/L(IR) ratio, again measured using the H$_2$O(2$_{02}$$\to$$1_{11}$) line, is $(4.3 \pm 1.6) \times 10^{-5}$, over a factor of 5 times brighter than mean value shown by the \cite{yang13} sample. \smg\ has the highest L(H$_2$0)/L(IR) ratio even amongst the lensed high-$z$ ULIRGs (which themselves show water emission more luminous than average). 


This water emission, combined with the likely high level of excitation present in \smg, is again consistent with the presence of a compact and extreme star forming region (see \citealt{gonzalezalfonso10}), though -- as with the $^{12}$CO emission discussed above -- the observed water excitation may be enhanced by a differential lensing effect.

\subsection{Far-IR fine structure lines}
\label{sec:lines}

Section \ref{sec:spire} describes our {\it Herschel}/SPIRE FTS observations of \smg. As discussed above, we fail to robustly detect any of the far-IR fine structure lines falling within the band.
Fig. \ref{fig:FIRLines} shows our derived upper limits (in units of L$_{\rm line}$/L$_{\rm FIR}$), compared to the typical range of values for starburst systems compiled by \cite{spinoglio12}. With the possible exception of [OI]63$\mu$m (see below), all our upper limits are consistent with \smg\ having line luminosity ratios comparable to other starburst and IR-luminous galaxies.

It has long been known that the ratio between far-IR line luminosity and far-IR continuum emission has a connection to the total IR luminosity of a galaxy, with luminous and ultra-luminous IR galaxies (LIRGS and ULIRGs) exhibiting `deficits' in their far-IR fine structure line ratios, relative to the values for secular, non-starburst galaxies (e.g., \citealt{luhman03}; \citealt{abel09}). One explanation for this result is the effect of the ionization parameter $U$ (defined as the ratio between hydrogen ionizing photons and hydrogen atoms; \citealt{abel05}). In this model, galaxies with enhanced ionization parameters have a greater fraction of their UV photons absorbed by interstellar dust, and thus fewer photons available to photoionize the gas (e.g., \citealt{abel09}). 


As discussed by \cite{graciacarpio11}, a good proxy for the ionization parameter $U$ is the ratio between far-IR luminosity and molecular gas mass, L$_{\rm FIR}$/M(H$_2$)  -- galaxies with enhanced values of L$_{\rm FIR}$/M(H$_2$) tend to have weaker far-IR fine structure lines, with a `transition' towards weaker lines happening at $> 80 \; {\rm L}_{\sun} {\rm M}_{\sun} ^{-1} $. We calculate \smg\ to have a L$_{\rm FIR}$/M(H$_2$) ratio of $(250 \pm 70) \; {\rm L}_{\sun} {\rm M}_{\sun} ^{-1}$. This value would number among the highest recorded in the compilation by \cite{graciacarpio11}, corresponding to the galaxies with the weakest far-IR fine structure lines. This value is high even for luminous DSFGs -- the compilation of 32 CO-observed galaxies presented in \cite{bothwell13} has a mean ($\pm$ standard deviation) L$_{\rm FIR}$/M(H$_2$) ratio of $(115 \pm 80) \; {\rm L}_{\sun} {\rm M}_{\sun} ^{-1}$. The extreme L$_{\rm FIR}$/M(H$_2$) ratio observed in \smg\ suggests an unusually high ionization parameter, 
which would result in extremely weak far-IR fine structure lines, consistent with the lack of detection of lines in our SPIRE spectrum.

Recently, some results have suggested that the [OI]63$\mu$m line ratio may be higher in some high-$z$ DSFGs than in local (U)LIRGs -- that is, the DSFGs may exhibit less of a [OI]63$\mu$m line deficit than nearby merging systems. \cite{coppin12} found that a sample of DSFGs in the LABOCA Extended Chandra Deep Field South exhibit [OI]63$\mu$m to far-IR ratios of approximately 0.5--1.5 percent -- an order of magnitude brighter than seen in local (U)LIRGs. Those authors argue that the relatively `enhanced' [OI]63$\mu$m line strength suggests that the star formation mechanism in their sample of DSFGs is more comparable to secular, disc galaxies (which typically do not display a [OI]63$\mu$m deficit). 

Our upper limit on the [OI]63$\mu$m line luminosity in \smg\ is sufficient to constrain the [OI]63$\mu$m to far-IR ratio to be significantly below the values recorded by \cite{coppin12}, and at the lower end of the range for local (U)LIRGs. 
According to the model described above, this would suggest that star formation is more compact in \smg\ than in the unlensed DSFGs in 
the \citet{coppin12} sample.\\

\begin{figure}[htb]
\epsscale{1.1}
\plotone{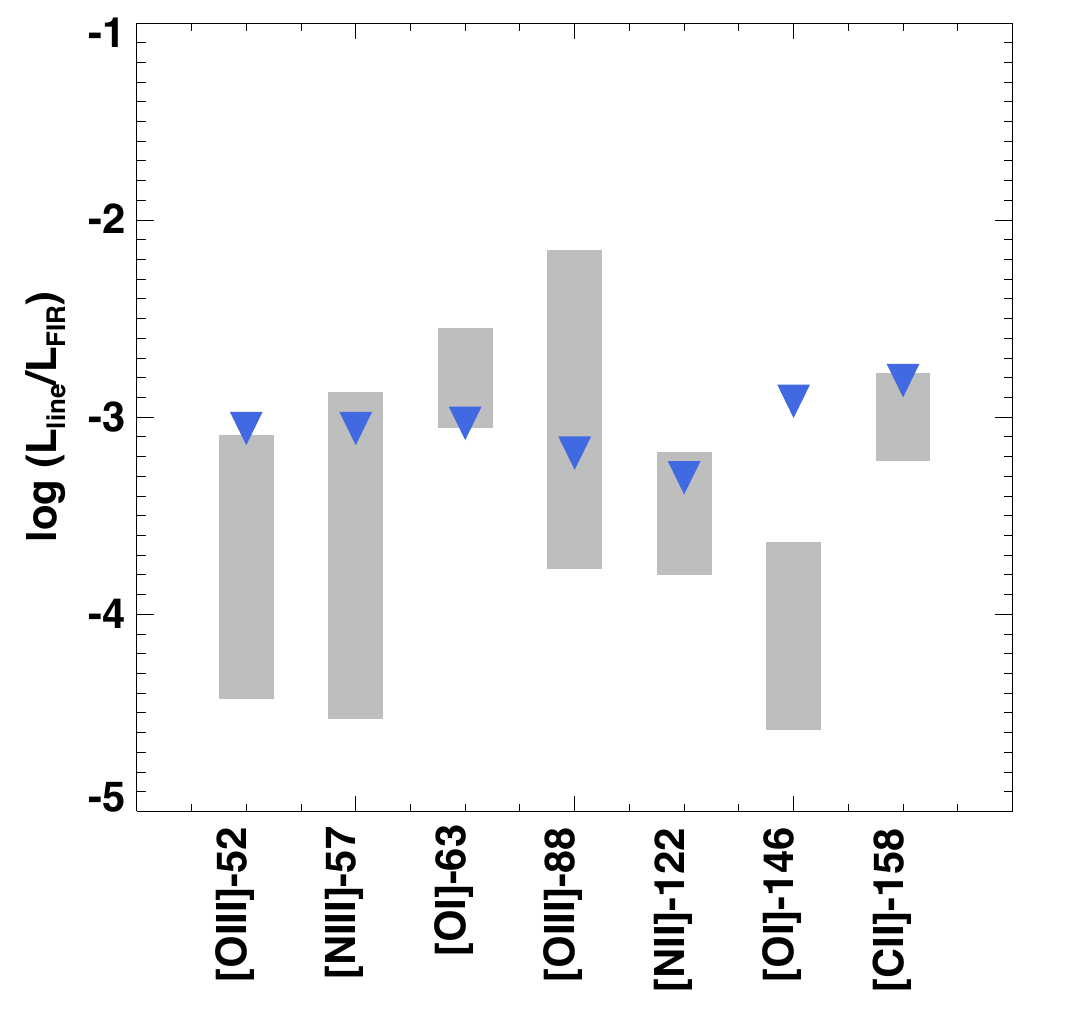}
\caption{Comparison of the FIR line upper limits with typical values for local (U)LIRGs. Blue triangles show the 2$\sigma$ upper limits on the FIR fine structure lines in \smg, derived using the SPIRE spectrum. Grey regions show the 1$\sigma$ range in values for (U)LIRG galaxies compiled by \citet{spinoglio12}.}
\label{fig:FIRLines}
\end{figure}




\section{Conclusions}\label{sec:conclusions}
We have presented a detailed observational study of a $z\sim2.8$ DSFG, which is being strongly gravitationally lensed by an intermediate-redshift elliptical galaxy. The lensing magnification allows us to carry out an array of multi-wavelength observations. A detailed lens model constructed from ALMA Band 7 data \citep{hezaveh13} reveals that \smg\ consists of two source-plane components (a bright compact component, and a dimmer diffuse component), and is magnified by a factor of $21\pm4$. Here we have used high-resolution {\it HST}/ACS imaging to build a model of the \smg\ lens galaxy. By convolving this with the {\it Spitzer}/IRAC PSF, we then used this model to de-blend the lower resolution IRAC imaging, separating the emission originating from \smg\ from that of the foreground lens.  \\


Our main results are as follows:

\begin{itemize}


\item{By SED fitting to a range of IR/mm data (including our de-blended IRAC fluxes), we find \smg\ to have a stellar mass of $(3.3 \pm 1.5) \times 10^{10} \; {\rm M}_{\sun}$, and a star formation rate of $760 \pm 100$ M$_{\sun}$/yr. This gives a sSFR of 25 Gyr$^{-1}$, placing \smg\ a factor of $\sim5$ above the main sequence of star-forming galaxies at this epoch, similar to the location (relative to the main sequence) of local merger-driven starbursts. Assuming a simple `closed box' situation, we calculate that \smg\ will exhaust its molecular gas supply in $\sim 20$ Myr.  }

\item{
We do not detect any far-IR fine structure lines in the spectrum of \smg. We attribute this to the presence of a high ionisation parameter $U$.}

\item{We detect several highly excited molecular emission lines (including H$_2$O and two high-$J$ lines of $^{12}$CO), which, when compared to other similar galaxies, suggest highly excited conditions in the ISM. This could be the signature of highly compact star formation, though boosting via differential lensing may also be contributing to this effect.  }

\end{itemize}
The derived physical properties all point towards \smg\ undergoing a highly compact, merger-driven mode of star formation, comparable to local ULIRGs, and unlike the extended mode of star formation discovered in some other DSFGs at this epoch. \\

{\it Facilities:} \facility{HST/WFC3}, \facility{Spitzer/IRAC},  \facility{ALMA},  \facility{APEX},  \facility{ATCA}, \facility{VLT:Kueyen}, \facility{SMA}, \facility{Herschel}

\acknowledgments

We thank the anonymous referee who provided comments which helped improve the clarity of this manuscript. The authors would like to thank N. Rangwala for sharing the M82 CO flux densities. MSB would like to acknowledge the hospitality of the Aspen Center for Physics where some of this manuscript was written. The Submillimeter Array is a joint project between the Smithsonian Astrophysical Observatory and the Academia Sinica Institute of Astronomy and Astrophysics and is funded by the Smithsonian Institution and the Academia Sinica. Support is provided by National Science Foundation grants AST-1009649, ANT-0638937, PHY-1125897, and PHYS-1066293. This paper makes use of the following ALMA data: ADS/JAO.ALMA \#2011.0.00957.S and \#2011.0.00958.S. ALMA is a partnership of ESO (representing its member states), NSF (USA) and NINS (Japan), together with NRC (Canada) and NSC and ASIAA (Taiwan), in cooperation with the Republic of Chile. The Joint ALMA Observatory is operated by ESO, AUI/NRAO and NAOJ. The National Radio Astronomy Observatory is a facility of the National Science Foundation operated under cooperative agreement by Associated Universities, Inc. Partial support for this work was provided by NASA through grant HST-GO-12659 from the Space Telescope Science Institute and awards for Herschel analysis issued by JPL/Caltech for OT1\_dmarrone\_1, OT1\_jvieira\_4, and OT2\_jvieira\_5.

\bibliography{../bibtex/spt_smg}{}

\appendix

\section{The lens galaxy}
\label{sec:lens}

Using the remaining IRAC flux (i.e., the flux residual after subtracting the background DSFG), we can also estimate a stellar mass for the foreground elliptical galaxy. \cite{Zhu10} give the following prescription for calculating the stellar mass of a galaxy from the IRAC $3.6 \mu$m luminosity:

\begin{equation}
 \log {\rm M}*  = (-0.79 \pm 0.03) + (1.19 \pm 0.01) \log \nu L_{\nu} [3.6\mu m]
\end{equation}

With the contribution from \smg\ subtracted, the foreground lens has a 3.6$\mu$m flux of $S_{3.6} = 382 \pm 65 \mu $Jy. 
We K-correct this flux to the rest frame 3.6$\mu$m using the prescription appropriate for early type galaxies given by \cite{Huang07}, which results in a (rest-frame) luminosity of $\nu L_{\nu} = (2.6 \pm 0.4) \times 10^{10} \, {\rm L}_{\sun}$. Using Eq. A1, this then gives a stellar mass of $(3.9 \pm 0.8) \times 10^{11} \, {\rm M}_{\sun}$.  

While the foreground lens galaxy of \smg\ is certainly a massive elliptical in a somewhat crowded environment, it is unlikely that it is the Brightest Cluster Galaxy (BCG) of a small cluster. As can be seen in from our HST/WFC3 imaging (see Fig. \ref{fig:ir}, top right panel), associated structure extends less than 100 kpc from the lensing galaxy, far smaller than the typical extent of even a small cluster. Combined with the relatively modest stellar mass (for a BCG), this suggests that \smg\ represents a galaxy-lensed system, rather than a cluster lens.

\section{Deconvolution without the ring-shaped normalising component}
\label{sec:ring}

As we discuss in  \S3.2.2 above, it is important to note that the `ring'-like structure of \smg\ seen in the IRAC residuals are not a result of using the normalising ring component (which is included only to achieve an optimum subtraction). Even crude normalization techniques (such as simply matching the peak flux of the model to the peak flux of the real image) recover ring-shaped residuals that match the ALMA sub-mm continuum. 

These cruder methods do not find an optimum subtraction (due to the DSFG emission being spread across the lens galaxy by the IRAC PSF), but it is important to stress that the ring-like residuals do not result from the use of the ring-shaped normalization component. 

Fig. \ref{fig:noring} shows a comparison of the 3.6 $\mu$m and 4.5 $\mu$m residuals obtained with and without the use of the dummy normalising component. While the quality of the subtraction is lower in the latter case (with some over-subtraction present in the center of the ring), we still easily recover the ring-like structure associated with \smg, demonstrating that it is real.

\begin{figure}[htb]
\epsscale{0.65}
\plotone{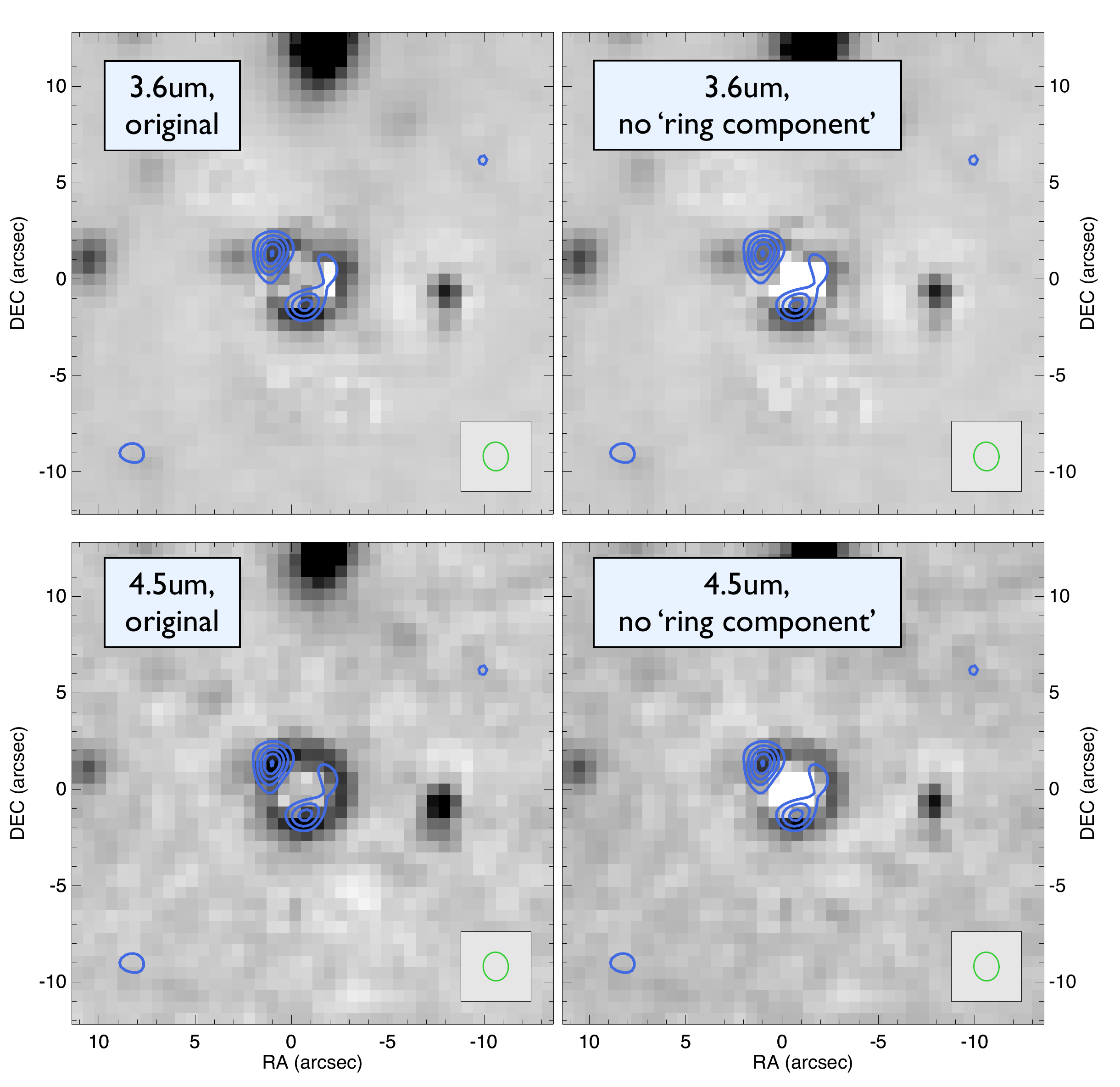}
\caption{A comparison of IRAC subtraction residuals (at both 3.6 $\mu$m and 4.5 $\mu$m), obtained with and without the use of a dummy normalising `ring' component. The image colorscale has been stretched in order to highlight `over-subtraction' issues. 
In both bands, the subtraction without the use of the ring component is poorer than our original, properly normalised data (presented in \S3.2.2 and Fig. 6), due to over-subtraction in the center of the rings. We include these comparison images to demonstrate that ring-like residuals (which match the ALMA emission) are not an artefact of the normalising component, but are present even when using more simplistic normalisation methods. }
\label{fig:noring}
\end{figure}

\end{document}